\documentclass[twocolumn]{pazhb_eng}

\usepackage{graphicx}
\usepackage{amsmath}
\usepackage{rotating}
\usepackage{lscape}
\usepackage{ifpdf}
\usepackage{breqn}
\usepackage[hyphens]{url}

\newcommand {\be}{\begin{equation}}
\newcommand {\ee}{\end{equation}}

\begin{document}
\journalinfo{2018}{44}{4}{248}[264]

\title{Determining the Absolute Magnitudes of Galactic-Bulge Red Clump Giants in the Z and Y Filters of the Vista Sky Surveys and the IRAC Filters of the Spitzer Sky Surveys }
\author{\bf D.I. Karasev\email{dkarasev@iki.rssi.ru}, A.A.Lutovinov\\
\it{Space Research Institute, Moscow, Russia\\
}}
\shortauthor{}
\shorttitle{}
\submitted{25 April 2017}

\begin{abstract}
The properties of red clump giants in the central regions of the Galactic bulge are investigated in the photometric $Z$ and $Y$ bands of the infrared VVV (VISTA/ESO) survey and the [3.6],
[4.5], [5.8], and [8.0] bands of the GLIMPSE (Spitzer/IRAC) Galactic plane survey. The absolute magnitudes for objects of this class have been determined in these bands for the first time: $M_{Z}= -0.20\pm0.04$ and
$M_{Y}=-0.470\pm0.045$, $M_{[3.6]}= -1.70\pm0.03$, $M_{[4.5]}= -1.60\pm0.03$,
$M_{[5.8]}= -1.67\pm0.03$ and $M_{[8.0]}= -1.70\pm0.03$. A comparison of the measured magnitudes with the predictions of theoretical models for the spectra of the objects under study has demonstrated good mutual agreement and has allowed some important constraints to be obtained for the properties of bulge red clump giants. In particular, a comparison with evolutionary tracks has shown that we are dealing predominantly with the high- metallicity subgroup of bulge red clump giants. Their metallicity is slightly higher than has been thought previously, $[M/H]\simeq0.40$ ($Z \simeq 0.038$) with an error of $[M/H]\simeq 0.1$ dex, while the effective temperature is $4250\pm150$\,K. Stars with an age of 9-10 Gyr are shown to dominate among the red clump giants, although some number of younger objects with an age of $\sim8$ can also be present. In addition, the distances to several Galactic bulge regions have been measured, as D = 8200-8500 pc, and the extinction law in these directions is shown to differ noticeably from the standard one.

\medskip
\end{abstract}

\section{INTRODUCTION}

At present, the most commonly used filters in
infrared astronomy are $J, H$ and $Ks$. Owing to the
success of the 2MASS all-sky survey, they are the
"golden standard" for the near-infrared photometry.
Properties of objects of various classes, including red
clump giants (RCGs), have been studied extensively
in these filters (see, e.g., Paczynski and Stanek 1998;
Girardi 1999, 2016). In particular, the absolute Ks
magnitude for bulge RCGs $M_{Ks} = -1.61 \pm 0.03$ (Alves 2000) and the corresponding intrinsic color $(J-Ks)_0 = 0.68\pm 0.03$ (Gonzalez et al. 2012) are
adopted as the reference values. Despite the fact
that the RCG population have a more complex and
inhomogeneous structure than has been assumed
previously (Gontcharov 2008; Bensby et al. 2013;
Gontcharov and Bajkova 2013; Gonzalez et al. 2015),
using the red clump centroid as a major tool for
working with RCGs remains typical.

New sky surveys, in particular, VISTA (VVV and
VHS) and UKIRT, have appeared in recent years.
Apart from the photometry in the above three filters,
measurements in several new ones, in particular, in
the $Z$ and $Y$ filters that are already very close to the
optical band, have also been performed. Mid-infrared
Galactic surveys, for example, GLIMPSE onboard
the Spitzer observatory in the [3.6], [4.5], [5.8], and
[8.0] filters, are also of great importance. Using these data extends possibilities for investigating RCGs. In particular, photometric estimates
made in a large number of filters can improve our understanding
of the properties of these objects and can help in
estimating the metallicity variations in the Galactic
bulge. Furthermore, using RCGs, one can study
the interstellar medium, in particular, determine the
extinction value and law. Using additional filters leads
to a more accurate and correct understanding of its
properties.

However, before turning to the solution of particular problems, it is necessary to determine the reference points, namely the absolute magnitudes of RCGs in the corresponding filters. This is primarily
true for the filters in which these magnitudes are not
yet known -- the $Z$ and $Y$ filters of the infrared VVV
(VISTA/ESO) survey and the [3.6], [4.5], [5.8], and
[8.0] bands of the GLIMPSE (Spitzer/IRAC) survey.

In this paper we consider several indirect methods
to determine the absolute magnitudes of RCGs localized in the Galactic bulge. Before turning to their
description and the results, several words should be
said about the direct method. This method is the most
obvious way of obtaining the absolute magnitudes
and is based on a comparison of the catalogues from
the surveys under study with the catalogue of stars
with known parallaxes. In particular, this approach
was applied by Laney et al. (2012) to determine the
absolute magnitudes of nearby RCGs from 2MASS
and Hipparcos data. This is also justified, because the
absolute magnitudes of RCGs in the bulge and the
solar neighborhood for the infrared agree satisfactorily
between themselves.

The VISTA surveys and GAIA parallax measurements are a natural combination to determine the
absolute $Z$ and $Y$ magnitudes of RCGs. However,
this approach does not work at the current stage. One
of the reasons is as follows: RCGs in the absence
of a significant interstellar extinction are too bright
even at fairly large distances, and their magnitudes in
the VISTA surveys are among the overexposed ones.
For example, the observed $Ks$ magnitude of RCGs is
$\approx8.3$, $\approx10$, and $\approx11$ at distances of 1, 2, and 3 kpc,
respectively. At the same time, according to the data
description\footnote{\url{http://www.eso.org/rm/api/v1/public/releaseDescriptions/80}}, stars with magnitudes $m_{Ks}\simeq10-12$ are overexposed in the VVV/VISTA survey.

In contrast, using more distant stars is so far
impossible due to the insufficient accuracy of the
parallaxes measured for them by the GAIA observatory and the presence of several significant systematic effects (Astraatmadja and Bailer-Jones 2016;
Gontcharov 2017). Furthermore, at large distances
the influence of interstellar extinction is already difficult to neglect.
Under such conditions, the indirect methods of estimating the absolute magnitudes play an important role.

\section{INSTRUMENTS AND DATA}

In this paper we used the publicly available photo-
metric data from VVV/ESO\footnote{\url{https://vvvsurvey.org}} Data Release 4. 
For our study we selected the photometric magnitudes measured in the third standard $2\arcsec$. aperture. To perform our study in the mid-infrared, we used the combined mid-infrared data from the GLIMPSE\footnote{\url{http://irsa.ipac.caltech.edu/data/SPITZER/GLIMPSE/}} surveys of the Spitzer Space Telescope obtained with the IRAC instrument. Detailed characteristics of the
surveys and the filters can be found at these links, and
they are briefly described in the final table containing
our results.

To compare our results with theoretical stellar
evolution models, we used the isochrones from the \emph{PARSEC}\footnote{\url{http://stev.oapd.inaf.it/cgi-bin/cmd}} database (Bressan et al. 2012; Marigo et al. 2017). We also used the \emph{SYNPHOT}\footnote{\url{http://www.stsci.edu/institute/software_hardware/stsdas/synphot}} package for synthetic photometry and the \emph{Kurucz}\footnote{\url{http://www.stsci.edu/hst/observatory/crds/k93models.html}} atlas of
stellar atmospheres.

\section{DETERMINING THE ABSOLUTE MAGNITUDES OF RCGs IN THE Z AND Y FILTERS}

\subsection{The First Method}\label{sec:undir1}

Since direct estimations of the absolute magnitudes for nearby RCGs are fraught with a number of objective difficulties (see above), below we will investigate the properties of RCGs localized directly in
the Galactic bulge, in the vicinity of Baade's window.
This and next sections will be devoted to describing
the methods and elaborating the most optimal approach using, as an example, the determination of the absolute $Z$ and $Y$ magnitudes as the least-studied
ones.
 
First of all, note that bulge RCGs have been
comprehensively studied previously, and the general
approaches are well known (see, e.g., Paczynski
and Stanek 1998; Dutra et al. 2003; Revnivtsev
et al. 2010; Gontcharov and Bajkova 2013). This
section is based on the method proposed by Karasev
et al. (2015) to determine the local extinction by in-
vestigating the distribution of RCGs in narrow strips
perpendicular to the Galactic plane. The strip width
is $2\arcmin$, which allows the scatter of distances to RCGs
due to the variability of the bulge structure along
the Galactic longitude to be minimized (Nishiyama
et al. 2009; Gonzalez et al. 2012). The vertical strip
size spans the range of Galactic latitudes from $-5^\circ$ to $+5^\circ$ . Such a significant strip "height" is admissible,
because it has been shown in several studies (see,
e.g., Gerhard and Martinez-Valpuesta 2012) that
the distance to the bulge changes little along the
Galactic latitude. Thus, within such a strip it will
everywhere be approximately the same. In addition,
Gonzalez et al. (2015) concluded that the metallicity
also depends weakly on the Galactic latitude (for $|b|<5^\circ$), i.e., within such a strip it must remain constant. It should also be noted that in the subsequent
analysis, as in Revnivtsev et al. (2010) and Karasev et al. (2010a), all of the absorbing material is assumed to be in front of the bulge.

Figures 1 and 2 show examples of the color -- apparent magnitude diagrams for all stars of such a strip in different sets of filters. The branch formed by
RCGs from different parts of the strip (i.e., located at
different latitudes) is clearly seen on each diagram.
Such RCGs have different extinctions and, accordingly, different shifts in magnitude and color from the true, dereddened state (in accordance with the classic
formula $m=M+5log_{10}D -5 + A $, where $m$ -- is the
apparent/measured magnitude, $M$ -- is the absolute
magnitude, $D$ is the distance and $A$ is the extinction). Since the distance to all parts of the strip may be
considered to be the same (i.e., the distance modulus
in this formula shifts equally all points along the vertical axis), the slope of the RCG branch is determined
exclusively by the ratio of the total extinction in the
corresponding filter to the selective one, i.e., by what
is called the extinction law in the literature. Thus,
from the slope of the RCG branch we can determine
the extinction law for the chosen strip.

Note that investigating the giants in such strips
also allows the distance to the bulge in a chosen
direction to be estimated (see, e.g., Revnivtsev et al. 2010; Karasev et al. 2010b). Knowing the absolute magnitude of RCGs at least in two filters
is a necessary condition for calculating the distance
in this case. As has been noted above, we know
their absolute Ks magnitude ($M_{Ks}$), which was also
obtained for Baade's window (Alves 2000), and the
dereddened color $(J-Ks)_{0}$ (Gonzalez et al. 2012).

For our study we also chose a strip in the vicinity
of Baade’s window with a longitude $l\approx1.7^\circ$ along
which there is quite a wide scatter of the extinction
(this is preferable for the estimates using strips). To
determine the extinction laws ($law_{JK} =
A_{Ks}/E(J-Ks)$, $law_{ZK}=A_{Z}/E(Z-Ks)$, etc.) with a better
accuracy, we first divided the strip into $2\arcmin \times 2\arcmin$ cells
(the total number of such cells is 300) and localized
RCGs on the color-apparent magnitude diagrams
constructed for them (Fig. 3). The cell size was
chosen in such a way that it contained a sufficient
number of RCGs for their subsequent identification
on the color-apparent magnitude diagram and the
determination of the centroid position. The chosen
cell size was previously used successfully for the VVV
survey data by Gonzalez et al. (2012). To avoid possible inaccuracies in determining the RCG centroid
due to the presence of main-sequence stars on the diagram, we first identified only the red giant branch on
it. Then we supplemented the main Gaussian model
used in searching for the RCG centroid by a linear
component that takes into account the contribution
of red giants of other classes (giant branch).

It should be noted that a statistically significant
detection of RCGs in several filters cannot be simultaneously obtained for all strip cells. This is particularly true for the central, most absorbed parts of the
strip and for the regions at its edges, where the giants
are not so numerous. Nevertheless, the total number
of such cells in the subsequent analysis of different
strips was always significant ($\sim220-280$).

  \begin{figure*}[]
  \centering
\includegraphics[width=1.17\columnwidth,trim={1cm 0 1cm 0cm},clip]{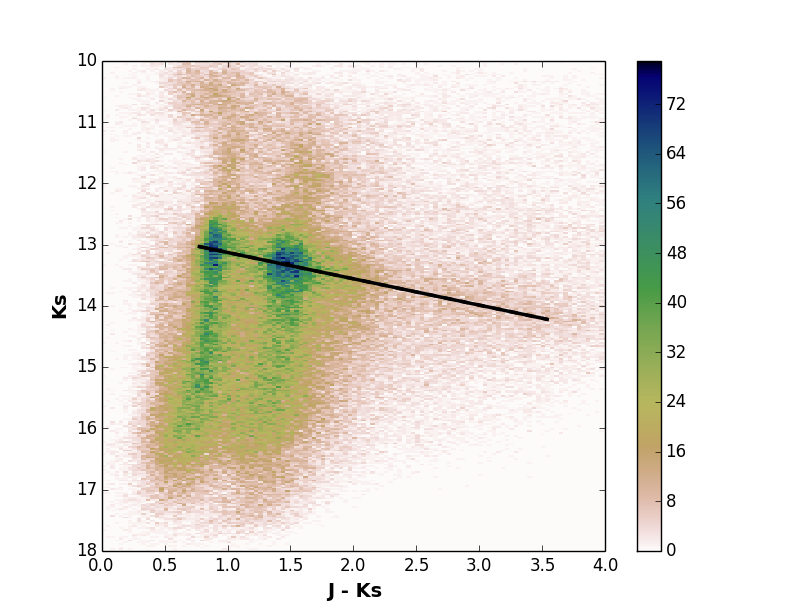}
\includegraphics[width=0.86\columnwidth,trim={2cm 6cm 1cm 1cm},clip]{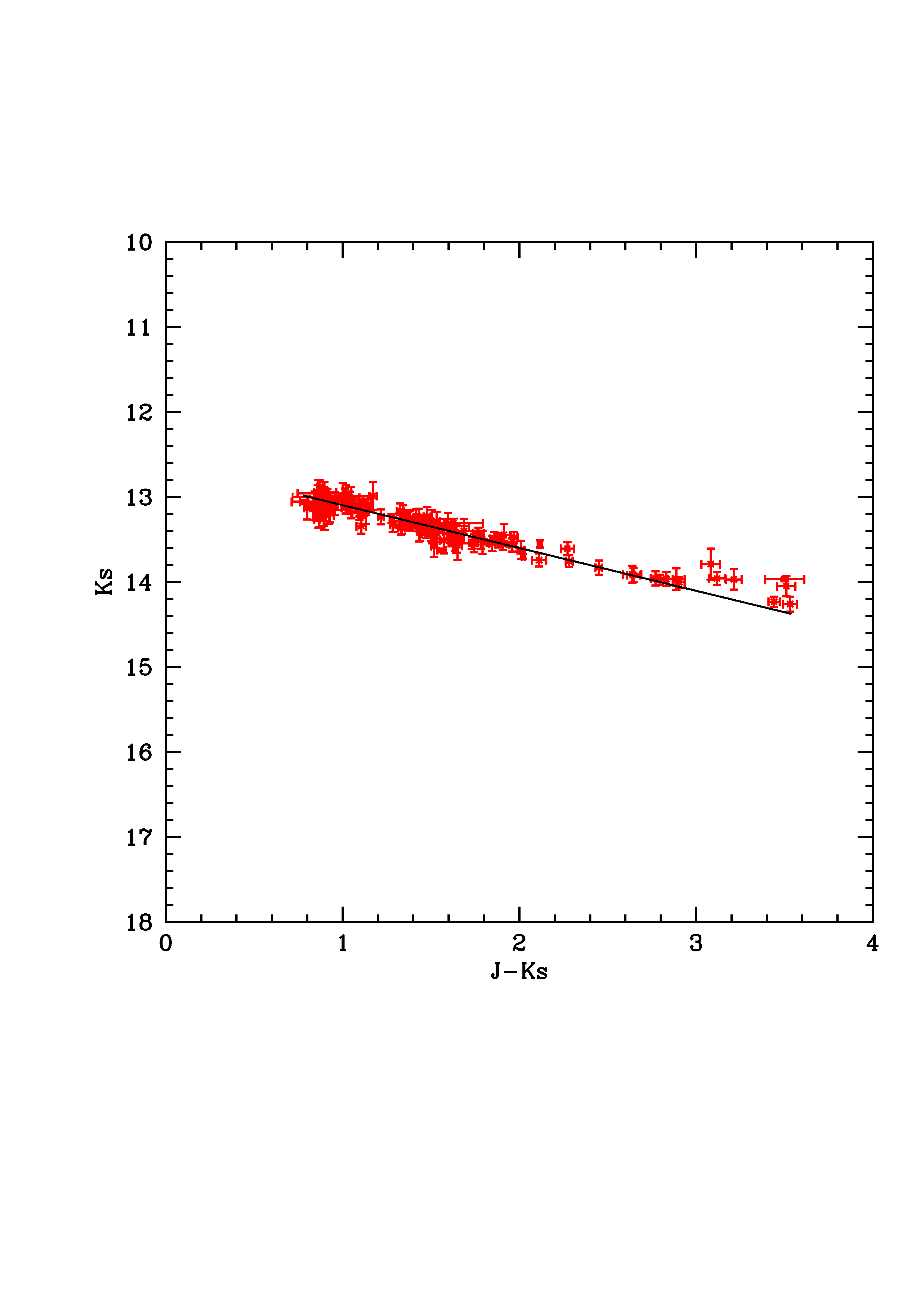}
\caption{(a) The color ($J-Ks$) --  magnitude ($Ks$), diagram constructed for all stars from the VVV survey in the strip with coordinates and a width of $2\arcmin$. The number of stars in the histogram bins is marked by the color
gradations on the scale rightward of the graph. (b) Observed magnitude of the RCG centroids localized in each of the $2\arcmin \times 2\arcmin$ strip regions versus color. The straight line on both panels indicates the linear model (1) that fits best the positions of the RCG centroids for the chosen sky strip.}
\label{fig:KJK}
\end{figure*}

  \begin{figure*}[]
  \centering
\includegraphics[width=1.17\columnwidth,trim={1cm 0 1cm 0cm},clip]{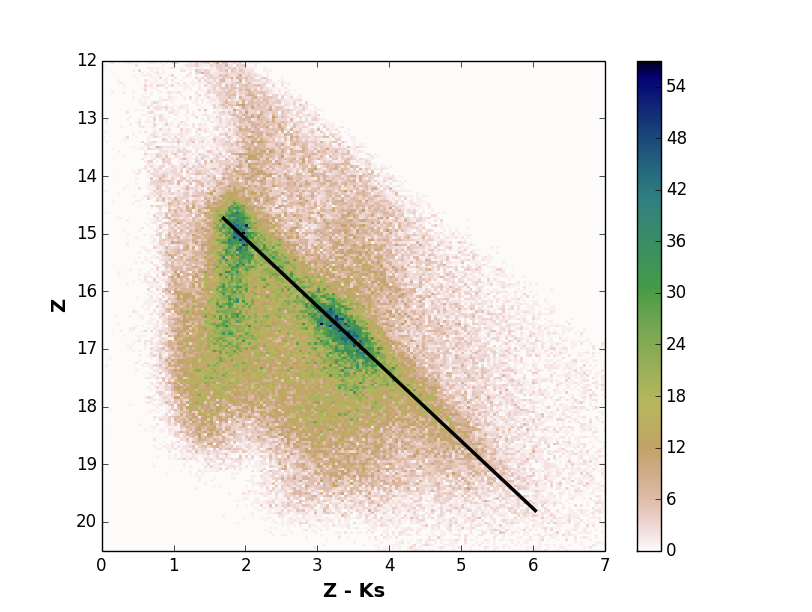}
\includegraphics[width=0.86\columnwidth,trim={2cm 6cm 1cm 1cm},clip]{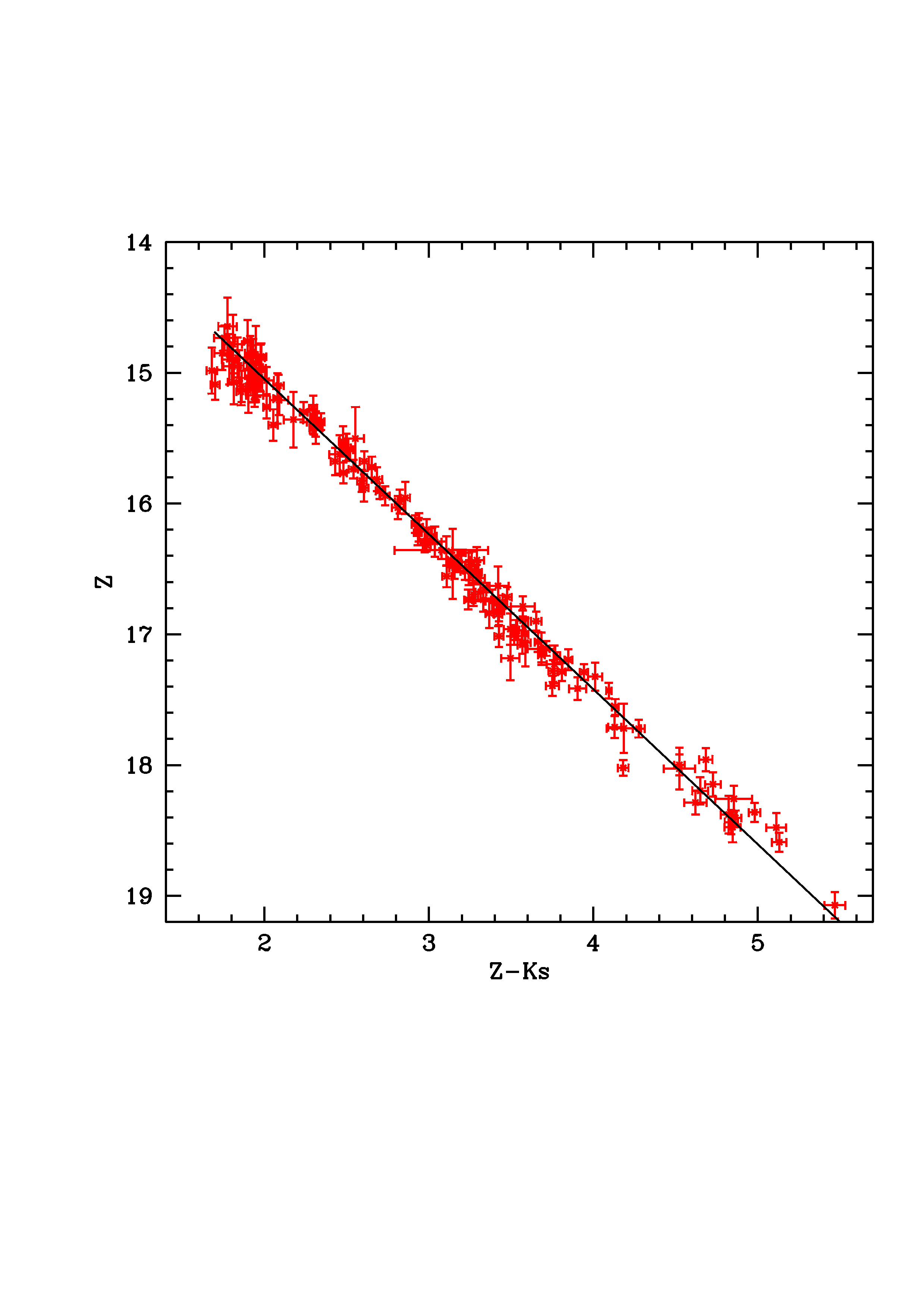}
\includegraphics[width=1.17\columnwidth,trim={1cm 0 1cm 0cm},clip]{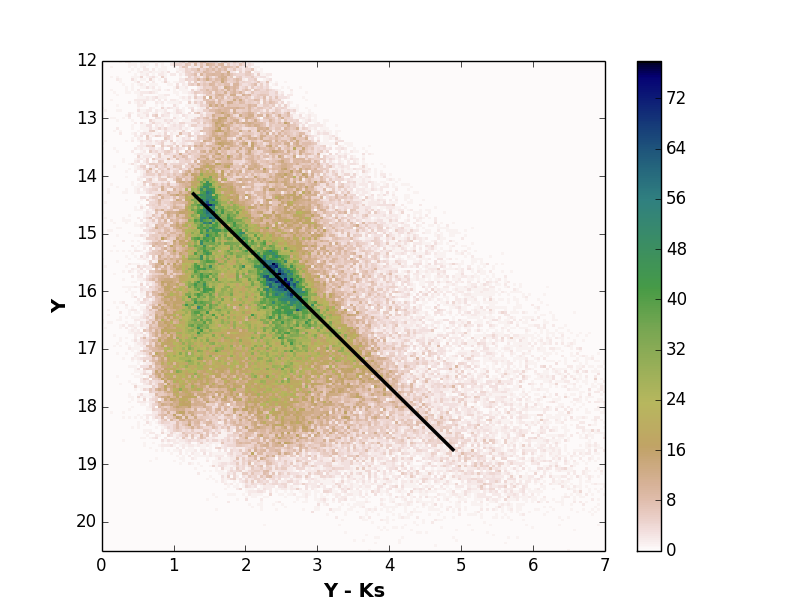}
\includegraphics[width=0.86\columnwidth,trim={2cm 6cm 1cm 1cm},clip]{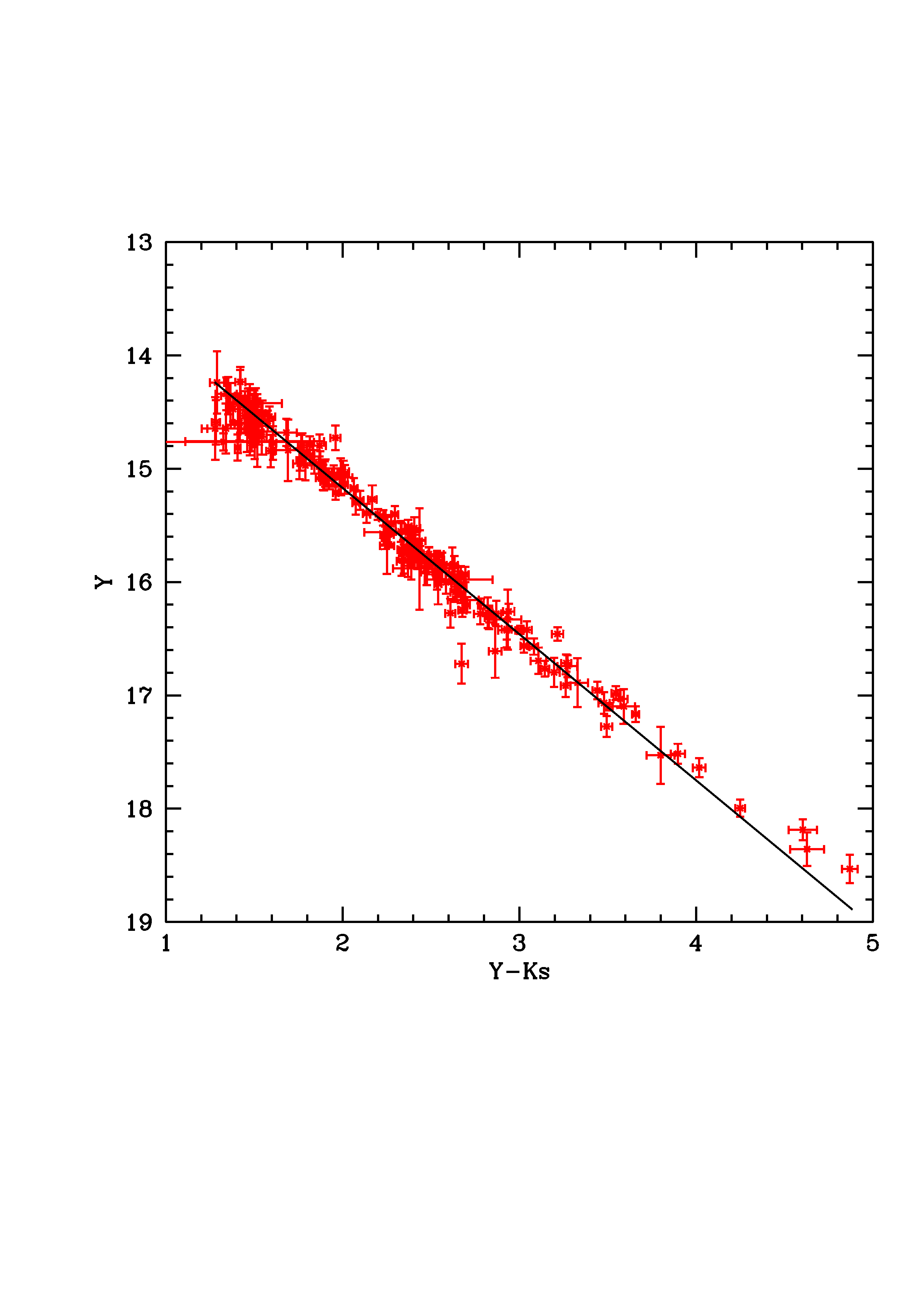}
\caption{Same as Fig. 1 for the ($Z, Ks$) and ($Y, Ks$) filters.  }  \label{fig:ZYK}
\end{figure*}

\begin{figure*}[]
\centering
\includegraphics[width=1\columnwidth,trim={1cm 6cm 0cm 1cm},clip]{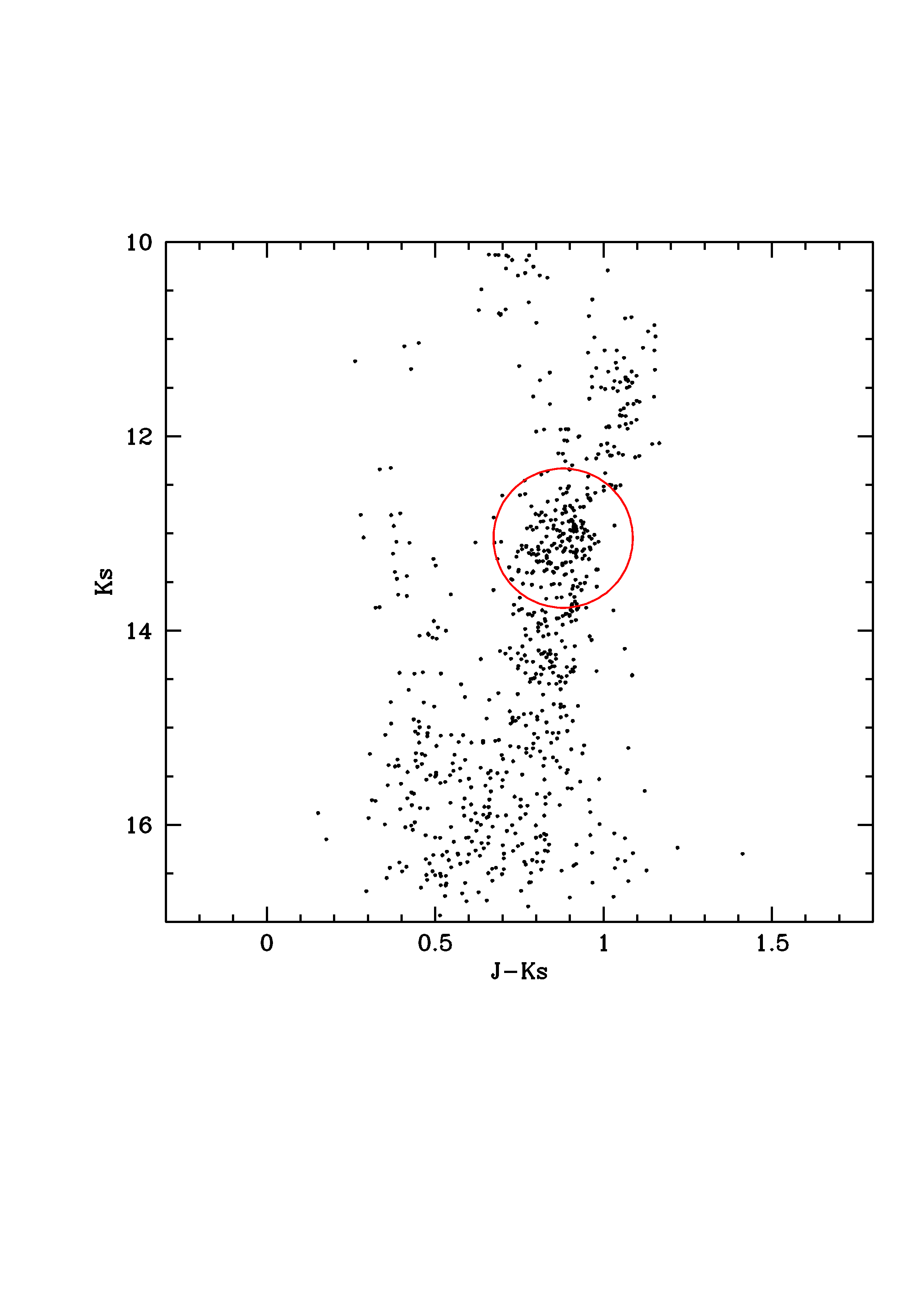}
\includegraphics[width=1\columnwidth,trim={1cm 6cm 0cm 1cm},clip]{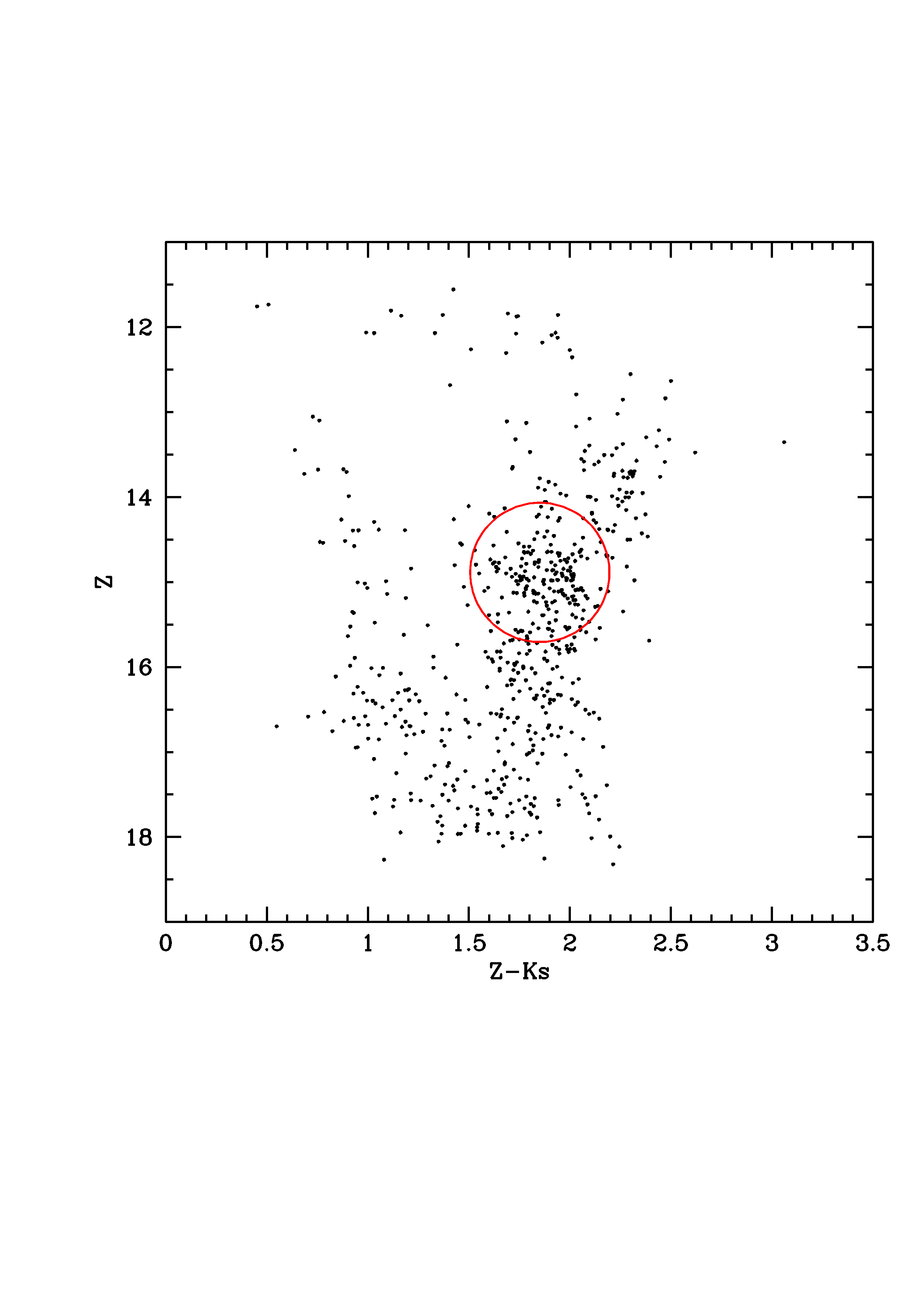}
\caption{An example of the color–apparent magnitude diagrams constructed in different filters for a $2\arcmin \times 2\arcmin$ cell of
the sky strip under study from the VVV survey data. The circles schematically mark the RCG positions.}
\label{fig:cell22}
\end{figure*}

Thus, the technique for estimating the absolute $Z$
and $Y$ magnitude can be formulated as follows.

1. By investigating the chosen sky strip in the $J$
and $Ks$ filters, for which the absolute magnitudes of
RCGs are known, we determine the distance to the
bulge $D$ (pc) in a chosen direction from the following relations:

\begin{equation}
m_{Ks}=law_{JK}\times(J-Ks)+const_{JK},
\end{equation}

\begin{dmath}
5- 5log_{10}D=M_{Ks}- law_{JK}\times(J-Ks)_0- const_{JK},
\end{dmath}

\noindent where $m_{Ks}$ is the observed Ks magnitude of the RCG
centroid, ($J-Ks$) is the measured color, and Eq. (1)
is the linear function that fits best the profile of the
RCG branch in the strip under study.

2. By fitting the color–magnitude relation for the
RCG centroids constructed in the ($Z, Ks$) and ($Y, Ks$) filters for all strip cells by a linear function, we obtain the corresponding slopes and constants.
Using the previously derived distance $D$, we find absolute $Z$
and $Y$ magnitudes of RCGs, $M_{Z}$ and $M_{Y}$:

\begin{dmath}
M_{Z} = (- law_{ZK}M_{Ks}+const_{ZK} + 5--5log_{10}D)/(1- law_{ZK}),
\end{dmath}
\begin{dmath}
M_{Y} = (- law_{YK}M_{Ks}+const_{YK} + 5--5log_{10}D)/(1- law_{YK}).
\end{dmath}

As a result, for the chosen strip we obtain
$m_{Ks}=0.43_{\pm0.02}(J-Ks)+12.70_{\pm0.02}$ (see Fig. 1), whence
the extinction law determined by its slope immediately
follows:
$law_{JK}=A_{Ks}/E(J-Ks)=A_{Ks}/(A_{J}-A_{Ks})=0.43\pm0.02$. It is important to
note that this law differs noticeably from the standard $A_{Ks} /E(J-Ks) \approx0.72 $ (Cardelli et al. 1989), which
is expectable for the bulge region and is consistent
with the conclusions of previous papers (see Table 1
and, e.g., Popowski 2000; Udalski 2003; Sumi 2004;
Nishiyama et al. 2009; Revnivtsev et al. 2010; Karasev et al. 2010a, 2010b; Nataf et al. 2016; Alonso-Garcia et al. 2017).

\begin{table*}
 \centering
  \footnotesize{
  \caption{Extinction laws obtained in this paper in comparison with the results of other authors}
  
  \begin{tabular}{c|c|c|c} 
  \hline
  Extinction
law	&   This paper     &      Cardelli et al. (1989)  & Alonso-Garcia et al. (2017)\\
 \hline
 $A_{Z}/E(Z-Ks) $   &    $1.17\pm0.02$	&	$1.3$  & $1.15$\\
 $A_{Y}/E(Y-Ks) $  & 	$1.23\pm0.01$	&	$1.44$	& $1.23$\\
  $A_{Ks}/E(J-Ks) $   &	$0.43\pm0.02$	&	$0.72$	&  $0.428$\\

\hline
\hline

\end{tabular}
}
\end{table*}

In addition, from the above relations we can esti-
mate the distance modulus $5- 5log_{10}D = -14.60 \pm 0.04 $ and the distance itself to the bulge in this part of the sky $D=8330\pm150$ pc. 

This result is important per se, and it agrees well with the present-day
estimates of the distances to the central bulge regions (see, e.g., Bhardwaj et al. (2017) and references therein). Note that if we take
$M_{Ks}=-1.63\pm0.03$, derived by Gontcharov (2017) for relatively nearby
stars as the reference absolute magnitude of RCGs,
then we will obtain the estimate for the distance to the bulge as 
$D=8410\pm150$ pc. It is consistent, within
the measurement uncertainties, with the above value.

Repeating the above steps 1 and 2 for the color--observed magnitude diagram in the
 ($Z, Ks$) and  ($Y, Ks$) filters, we obtain:
$m_{Z}=1.17_{\pm0.02}(Z-Ks)+12.75_{\pm0.03}$ and
$m_{Y}=1.23_{\pm0.01}(Y-Ks)+12.74_{\pm0.03}$  (Fig. 2),
and from Eqs. (3) and (4) we determine $M_{Z}= -0.18\pm0.43$ and $M_{Y}=-0.51\pm0.30$.

To test our results, we investigated another strip
farther from the Galactic center and located 12\arcmin\ away from the first one using the same technique. As a result, we obtained the following relations
and values:
$m_{Ks}=0.42_{\pm0.03}(J-Ks)+12.73_{\pm0.03}$,
$m_{Z}=1.17_{\pm0.02}(Z-Ks)+12.77_{\pm0.03}$ and
$m_{Y}=1.24_{\pm0.01}(Y-Ks)+12.74_{\pm0.02}$, whence from Eqs. (3)
and (4) we determine $M_{Z}= -0.16\pm0.45$ and $M_{Y}=-0.47\pm0.29$.

We see that the results of our measurements for
both strips agree between themselves much better
than they could, given the so significant measurement
errors.
To conclude this section, recall that the above
estimates were made by assuming the extinction law
in the infrared filters to be weakly variable along the
entire sky strip. The results of our measurements presented on the left panels of Figs. 1 and 2 and described
by a simple linear model, on the whole, confirm this
(the reduced value of $\chi^2 \approx1.2$). At the same time, it
is worth noting that on these diagrams there are also
points deviating significantly from the best-fit linear
models. Their presence can be explained both by the
existence of some local variations in the extinction
law, which is quite possible even in sufficiently narrow
strips (see, e.g., Gontcharov 2012; Nataf et al. 2013,
2016), and by the fact that the distance to the bulge
RCGs along the strip can have some variability, and
this variability will affect their observed magnitudes.

Thus, the proposed method yields the absolute
magnitudes of RCGs, but their uncertainties are too
large for these magnitudes to be used in subsequent
studies. At the same time, this method allowed us to
estimate the distance to the Galactic bulge quite accurately and to establish the extinction law (Table 1)
that, even despite its possible local variations, is much
better suited for the sky region under study than the
standard one.
  \begin{figure*}[]
  \centering
\includegraphics[width=1\columnwidth,trim={1cm 6cm 1cm 1cm},clip]{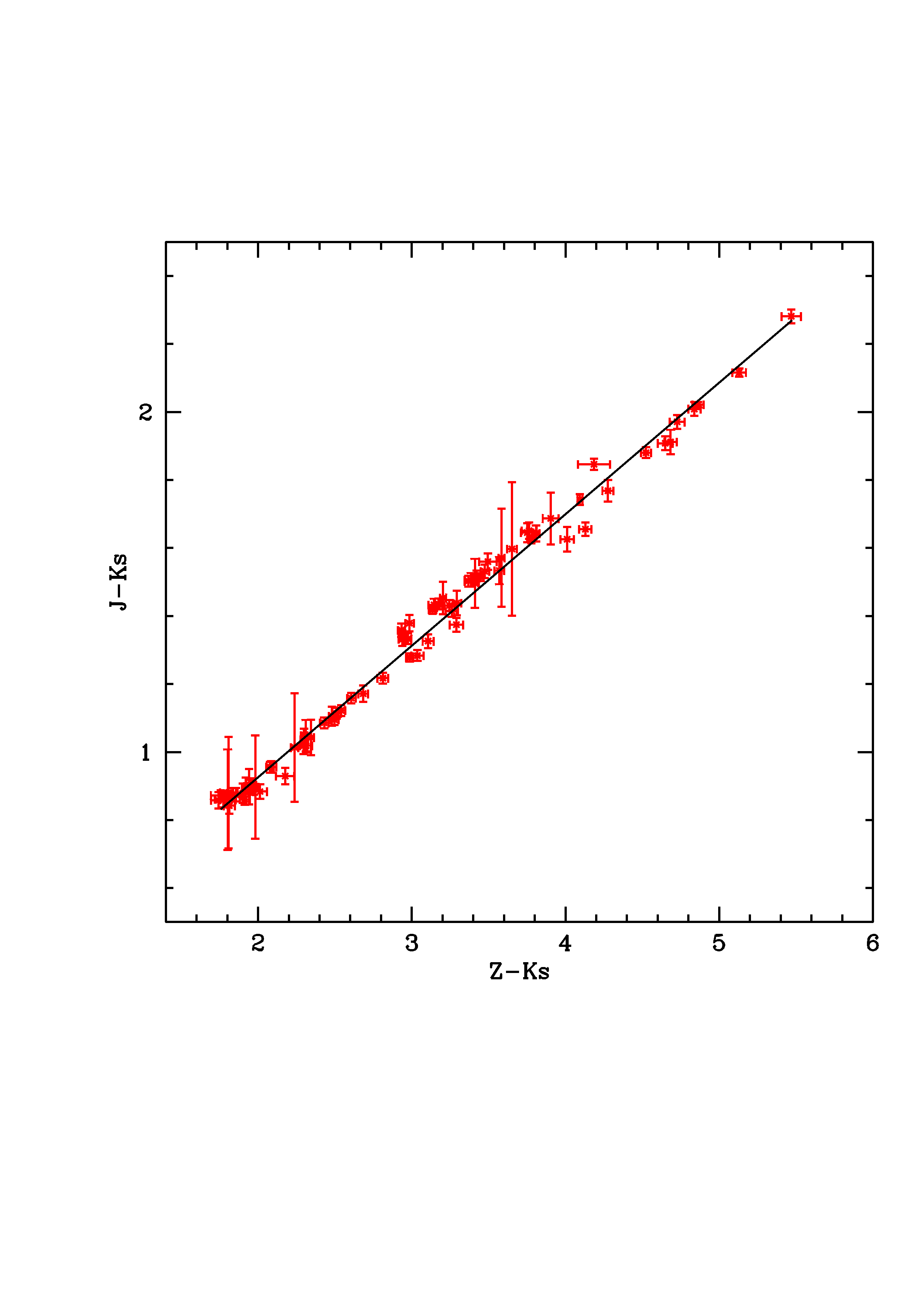}
\includegraphics[width=1\columnwidth,trim={1cm 6cm 1cm 1cm},clip]{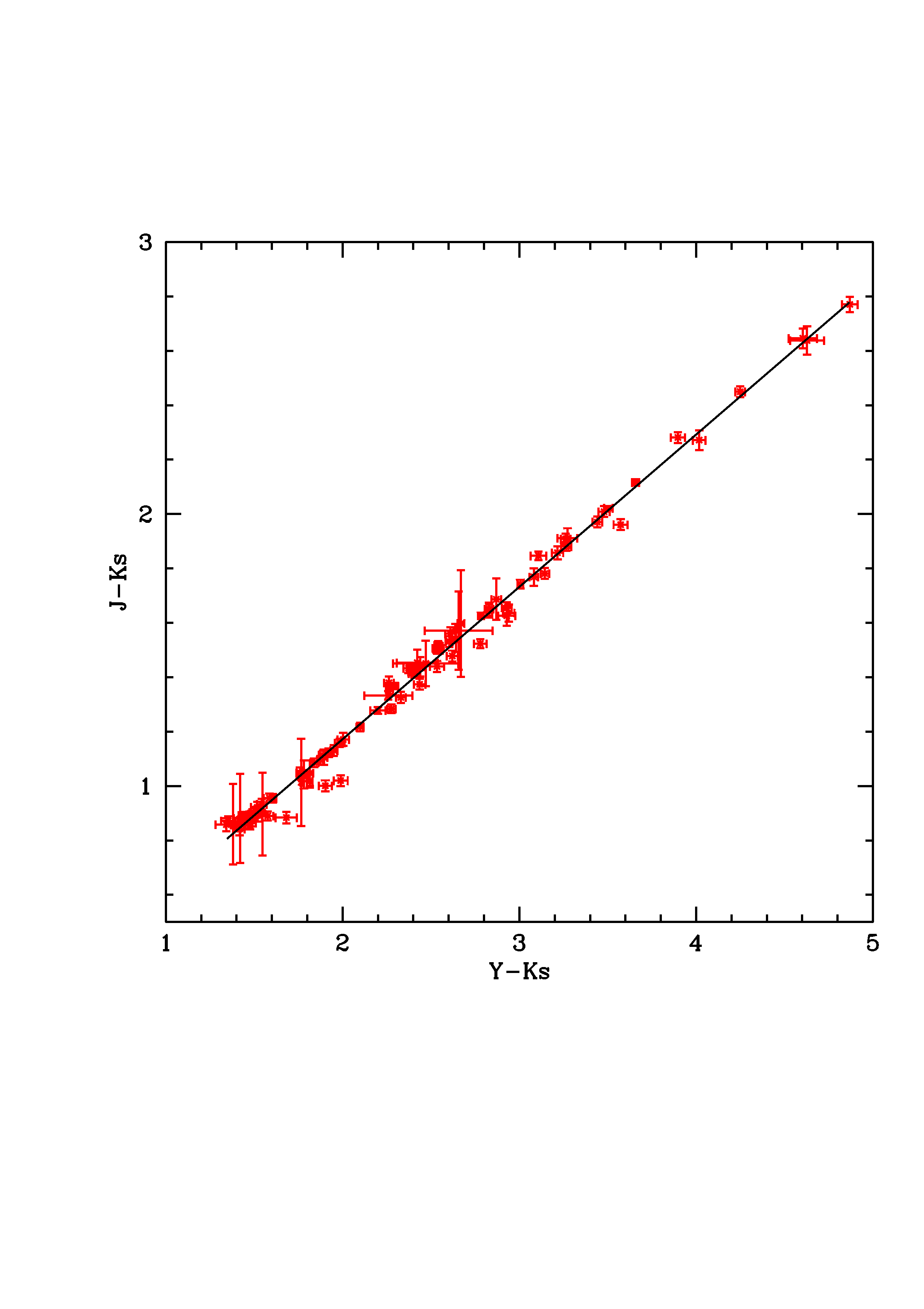}
\caption{(a) The color ($Z - Ks$) --  color ($J - Ks$)', diagram constructed for the strip with coordinates $l=1.7^\circ$, $b=[-5;5]$ from the VVV survey data.   (b) The color ($Y - Ks$) --  color ($J - Ks$), diagram constructed for the same sky
region. The straight lines on both panels indicate the linear models that fit best the corresponding dependences.}
\label{fig:COLORCOLOR}
\end{figure*}

\subsection{The Second Method}

We can attempt to reduce the measurement uncertainty of the absolute magnitudes determined in
the previous section by combining and averaging the
results for a large number of strips. However, in this
approach we inevitably introduce an additional inaccuracy that is also related to the possible metallicity
variations in the bulge (Gonzalez et al. 2015). If,
however, we somehow specially select the strips, then
the selection effect can manifest itself. Therefore, the
approach that allows the absolute magnitudes to be
reliably measured using only one strip is preferable.

Another way of reducing the final uncertainties is
to use exclusively the observed colors of the RCG
centroids, i.e., to use the $(J-Ks)-(Z-Ks)$ and
$(J-Ks)-(Y-Ks)$ color--color diagrams (see Fig. 4) instead of the color--observed magnitude diagrams.

As a result of this approach, we obtain two
sets of points that can be fitted by linear functions, $(J-Ks)=0.39_{\pm0.01}(Z-Ks)+0.151_{\pm0.015}$ and $(J-Ks)=0.56_{\pm0.01}(Y-Ks)+0.05_{\pm0.01}$. Given
the intrinsic color  $(J-Ks)_0$, we can determine the
intrinsic colors $(Z-Ks)_0$ and $(Y-Ks)_0$ and the absolute magnitudes of RCGs as $M_{Z}=-0.25\pm0.09$ and $M_{Y}=-0.485\pm0.06$. These values are consistent with the results of the previous method but excel them noticeably in accuracy.

Note that abandoning the use of the observed
magnitude of RCGs is also justified in that the natural
scatter of RCG magnitudes is noticeably larger than
the natural scatter of their colors (see below) and,
therefore, it is more difficult to correctly determine the
observed magnitude of RCGs than their color.

\subsection{The Third Method}

In the previous section we showed that abandoning the use of the observed magnitudes improved
significantly the situation with errors in the derived
absolute magnitudes, but they still remain fairly large.
Furthermore, both methods described above suggest
the constancy of the extinction law along the entire
strip, but, as has already been noted above, the presence of its, even if fairly small, variations must not be
ruled out completely. In this section we propose a
method of indirectly estimating the absolute magnitudes in which only the colors of RCGs in the cell are
used, but the assumption about the constancy of the
extinction law along the entire strip is not required.

To determine the absolute $Z$ magnitude of RCGs
($M_{Z}$) by this method, only the strip cells in which
the colors $(J-Ks)$ and $(Z-Ks)$ of RCGs were
simultaneously measured are suitable for us. To
determine $M_{Y}$, the cells in which the colors $(J-Ks)$ and $(Y-Ks)$ were simultaneously measured
are suitable. Below we describe the algorithm for
estimating the absolute magnitude in the $Z$ filter, but
the same approach is also valid for other filters.

\begin{figure*}[]
\centering
\includegraphics[width=1.02\columnwidth,trim={1.1cm 1 2cm 1cm},clip]{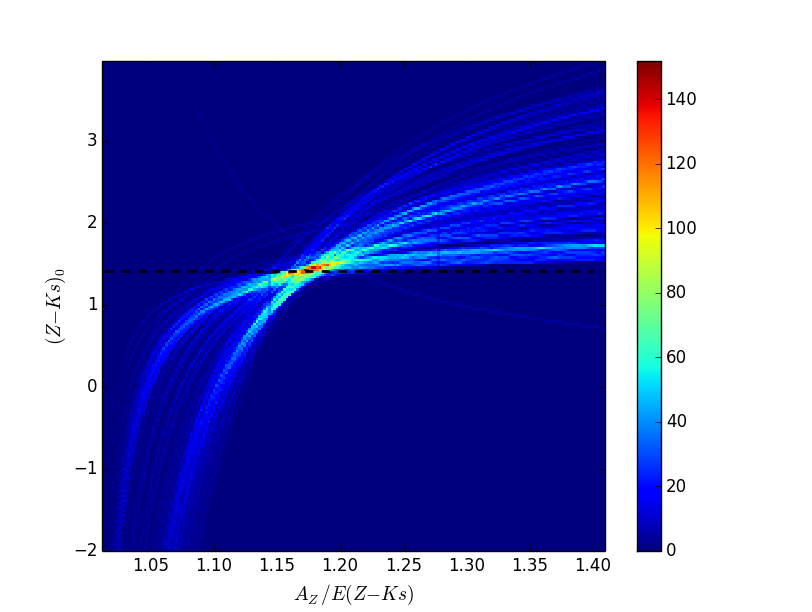}
\includegraphics[width=1.02\columnwidth,trim={1.1cm 1 2cm 1cm},clip]{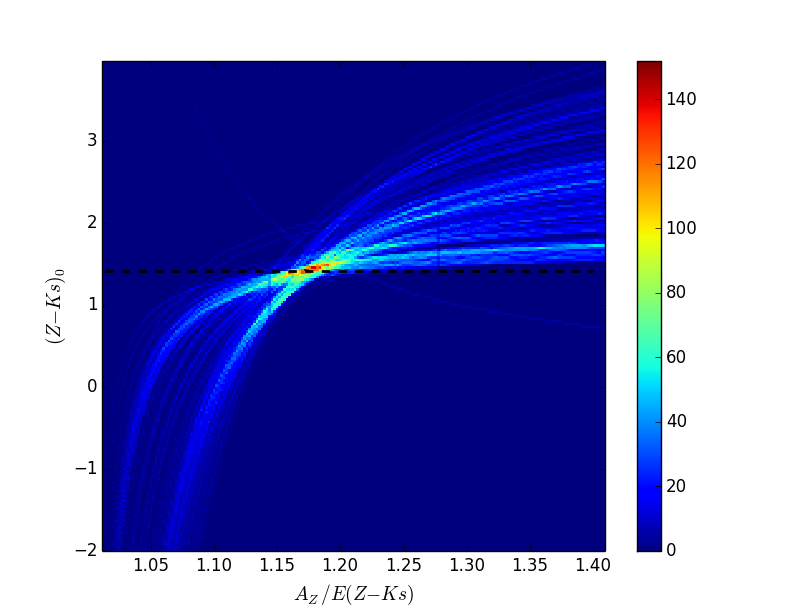}
\caption{Intrinsic color -- extinction law diagram. The region with the greatest density of points (the largest number of intersections between the tracks of different cells) corresponds to the true value of the corresponding color (for more details, see the text). (a) For the extinction law
$law_{JK}=A_Ks/E(J-Ks)=0.43$. (b) For the extinction law $law_{JK}=0.5$.
The dashed line marks the ordinate of the point of intersection between the tracks. The color gradation reflects the density of points
on the diagram. 
} \label{fig:method3_trek}
\end{figure*}

The observed color of RCGs depends on the intrinsic color and the relation between the extinctions:

\begin{equation}
 (Z-Ks)=(Z-Ks)_0 + (A_{Z}-A_{Ks}),
\end{equation}
    whence
\begin{equation}
 (Z- Ks)_0=(Z-Ks) - A_{Z}+A_{Ks},
\end{equation}

Thus, if we manage to determine the intrinsic colors
of the RCG centroids in individual cells $ (Z- Ks)_0=M_{Z}-M_{Ks}$ and $(Y- Ks)_0=M_{Y}-M_{Ks}$, then we
will find the sought-for absolute magnitudes, because
$M_{Ks}$ is known (see above). Obviously, the derived
intrinsic colors in different cells must be the same (i.e.,
they must coincide, within the measurement uncertainties), because everywhere we are dealing with the
properties of the same class of objects. Thus, $(Z-Ks)$ is a measurable quantity, the extinctions $A_{Z}$ and $A_{Ks}$ are related via the extinction law $law_{ZK}$ , and it only remains to determine $A_{Ks}$. This can be done
by using the fact that we know the color $(J-Ks)$ of
the RCG centroid in the same cell and the following
relation:

\begin{equation}
(J-Ks)=(J-Ks)_0 + (A_{J}-A_{Ks}),
\end{equation}

\noindent where we know $(J-Ks)_0$, while $A_{J}$ and $A_{Ks}$ are
related by the extinction law $law_{JK} = A_{Ks}/(A_{J}-A_{Ks})$.

Combining Eqs. (6) and (7), we find
\begin{dmath}
(Z- Ks)_0=(Z-Ks)- \frac{law_{JK}}{(law_{ZK}-1)}\times\times((J-Ks)-(J-Ks)_0),
\end{dmath}

It can be seen from the derived formula that the entire
dependence on extinction was reduced to the dependence on two extinction laws.

Let us now trace precisely how any variations in
the extinction laws affect the derived absolute magnitude (or intrinsic color) in different cells. For this
purpose, we will construct the true color -- extinction
law diagram (Fig. 5), where $law_{ZK}$ is along the horizontal axis and the intrinsic color $(Z-Ks)_0$ is along
the vertical axis. To understand the dependence on
the law $law_{JKs}$ , we will take two values: 0.43, i.e.,
the best-fit value from the previous section, and the
arbitrary one 0.5. It can be seen from the figure that,
depending on the extinction law $law_{ZK}$ , each of the
strip cells forms its personal track on such a diagram,
while changing the law $law_{JKs}$ simply shifted the
entire picture of tracks along the horizontal axis.

It is clear from the diagram that the derived intrinsic colors must be reduced to some solution suitable
for most of the cells. This solution is at the point (or
the points, if there are RCGs with different absolute
magnitudes in the chosen sky region) of intersection
between the tracks, and the ordinate of the point of
intersection will correspond to the sought-for intrinsic color. Moreover, it can be clearly seen from Fig. 5
that changing $law_{JKs}$ kept the ordinate of the point
of intersection unchanged. Thus, if several extinction
laws are present in the strip under study, then at
an invariable absolute magnitude this will give rise
to several points of intersection between the tracks
with equal ordinates spaced along the horizontal axis
apart.

Writing Eq. (8) for any two cells $i$ and $j$, we can
show that the intrinsic color of RCGs (or, in other
words, the projection of the point of intersection onto
the vertical axis) does not depend on any of the extinction laws:

\begin{dmath}
 (Z-Ks)_0=(Z-Ks)_i-  \frac{law_{JK}}{(law_{ZK}-1)}\times  \times((J-Ks)_i-(J-Ks)_0),
 \end{dmath}
 \begin{dmath}
 (Z-Ks)_0=(Z-Ks)_j-  \frac{law_{JK}}{(law_{ZK}-1)}\times  \times((J-Ks)_j-(J-Ks)_0),
 \end{dmath}
 whence:
 \begin{equation}
 \frac{law_{JK}}{(law_{ZK}-1)}=\frac{(Z-Ks)_j -(Z-Ks)_i}{(J-Ks)_j -(J-Ks)_i},
 \end{equation}
 and, finally,
 \begin{dmath}
 (Z-Ks)_0=(Z-Ks)_i- \frac{(Z-Ks)_j -(Z-Ks)_i}{(J-Ks)_j -(J-Ks)_i}\times((J-Ks)_i-(J-Ks)_0).
\end{dmath}

Note that in contrast to the previous methods,
where obtaining the result required assuming the extinction law to be invariable for the entire strip, here it
is actually necessary to assume the equality of the extinction laws simultaneously only for two cells. Using
Eqs. (9) -- (12), we can now link all of the cells from the
strip under consideration between themselves in pairs
and find the sets of possible intrinsic colors of RCGs
by investigating the density distribution of the points
of intersection along the $(Z-Ks)_0$ axis.


\begin{figure}[]
\centering
\includegraphics[width=1\columnwidth,trim={0cm 6cm 1cm 1cm},clip]{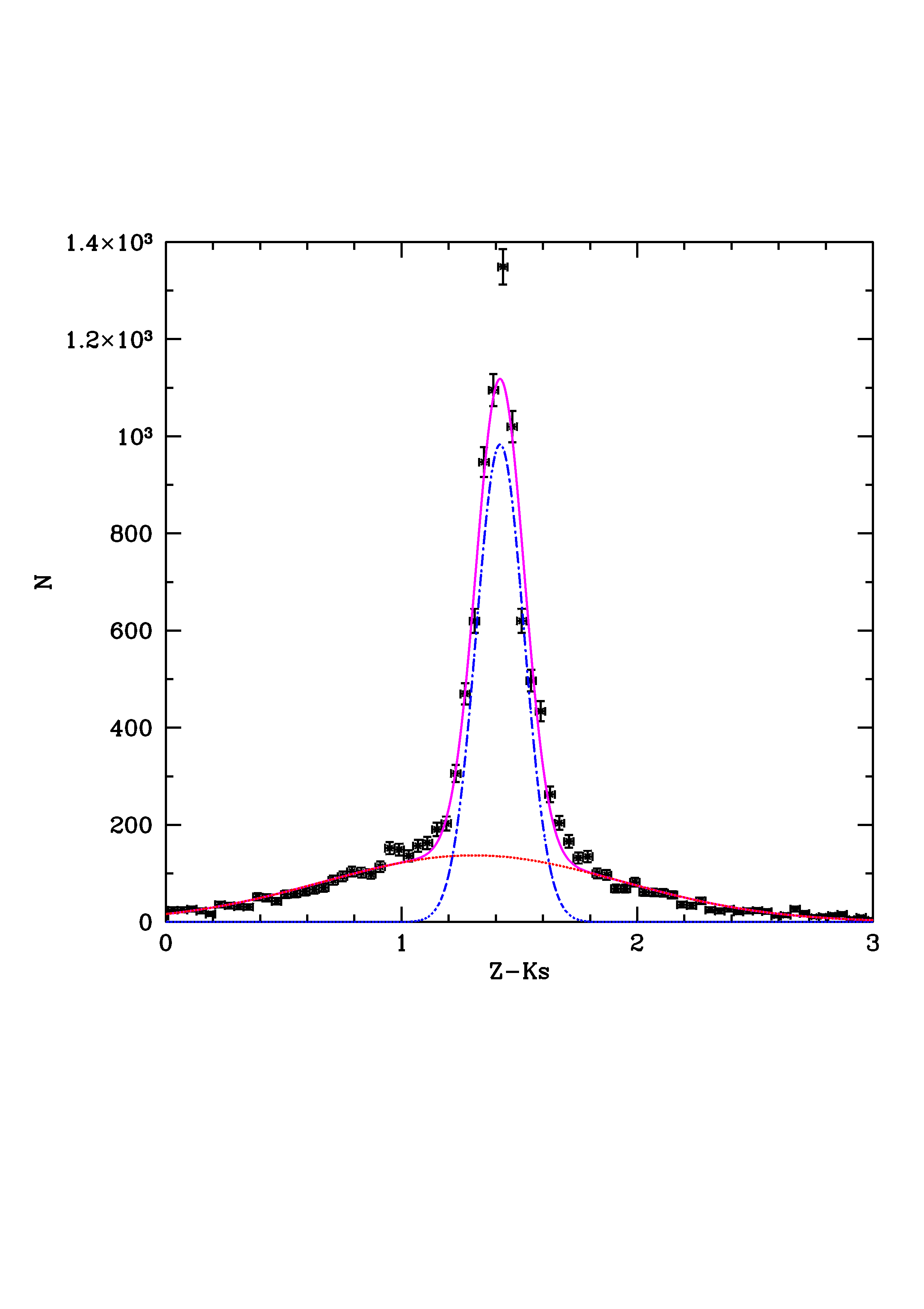}
\caption{The distribution of dereddened colors $(Z-Ks)_0$. The lines indicate the model that fits best the derived distribution (solid magenta line) and its components (dashed blue and dotted red lines).}
\label{fig:SINGLEZZK}

\end{figure}


The result of this approach is presented in Fig.\,6.
The only peak that corresponds to the intrinsic color
of RCGs in the corresponding filters, which can be
fitted by a Gaussian, is clearly seen. Besides, some
broadening at the base of the distribution, which can
be associated both with the statistical scatter of centroid points and with some variability of the extinction
law in the strip, is noticeable. This broadening can be
taken into account by adding an additional Gaussian
component to the model. Note that if there were
a sufficient number of RCGs with different absolute
magnitudes in the strip, then several significant peaks
would appear on the graph. Since this is not observed
in our case, we can argue that RCGs of one type with
the same (within the error limits) intrinsic color and
absolute magnitude dominate in the sky region under
study.
 
It is important to note that the derived color $(Z-Ks)_0$ is only one of the possible realizations, all of the
measured quantities in Eq.(12) have their own measurement errors. Therefore, to determine the color $(Z-Ks)_0$ and its measurement error more properly,
we used the so-called statistical bootstrap method. In
this method each of the measured values is selected
within the limits of its measurement error, whereupon
Eqs. (9) -- (12) are solved again, the derived distribution of $(Z-Ks)_0$ is fitted by a Gaussian, and its centroid is found. As a result of repeating such an analysis $\sim30000$ times, we obtained a sample of $(Z-Ks)_0$
whose mean and dispersion determine the
intrinsic color of RCGs and its measurement uncertainty, $(Z-Ks)_0= 1.414\pm0.025$ (Fig. 7). Note that
the statistical bootstrap method is used quite widely
in various problems of astrophysics, for example, to
properly determine the measurement errors of the periods of X-ray pulsars (for a more detailed description, see Boldin et al. 2013).

\begin{figure*}[]
\includegraphics[width=1\columnwidth,trim={0cm 6cm 0cm 1cm},clip]{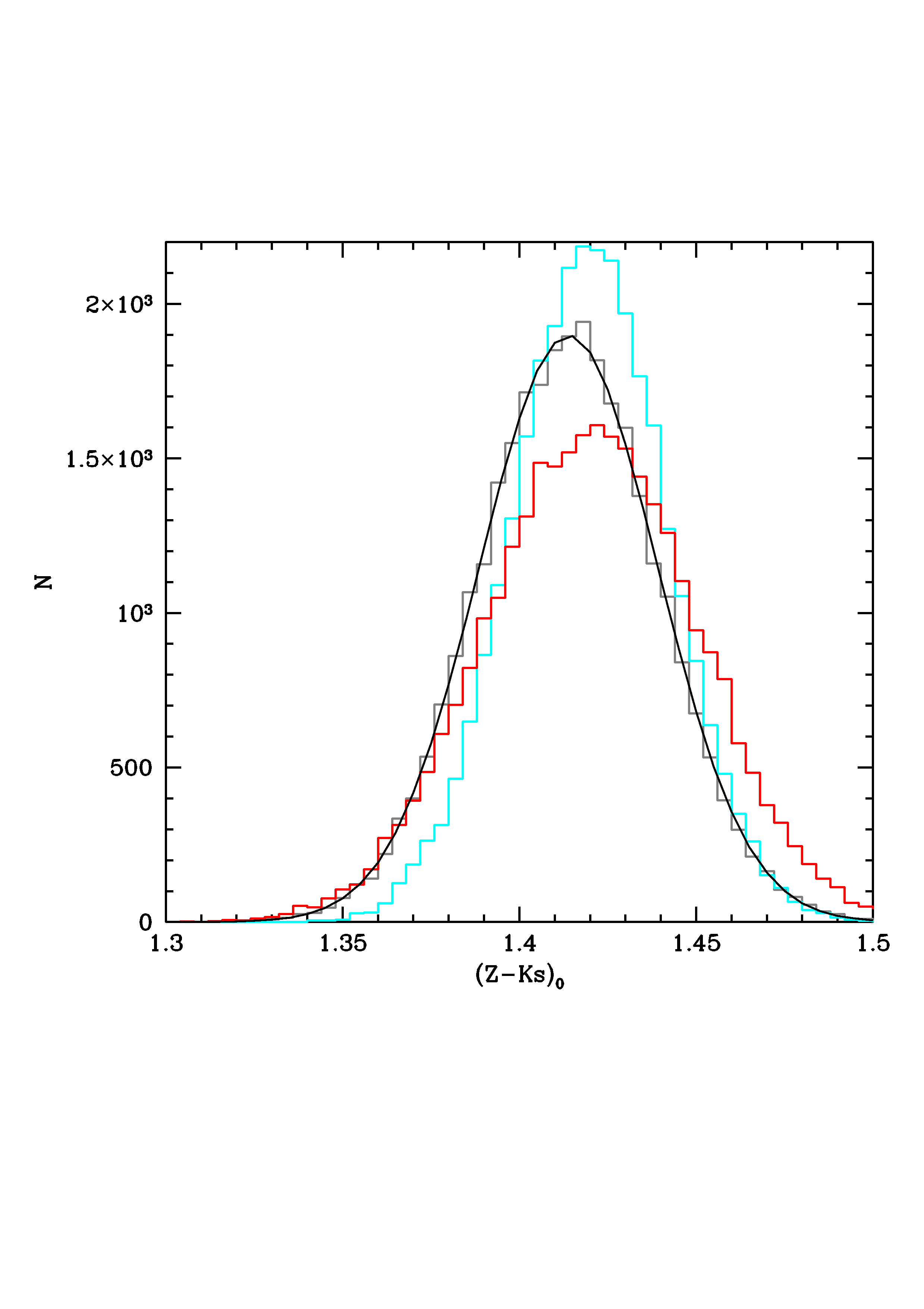}
\includegraphics[width=1\columnwidth,trim={0cm 6cm 0cm 1cm},clip]{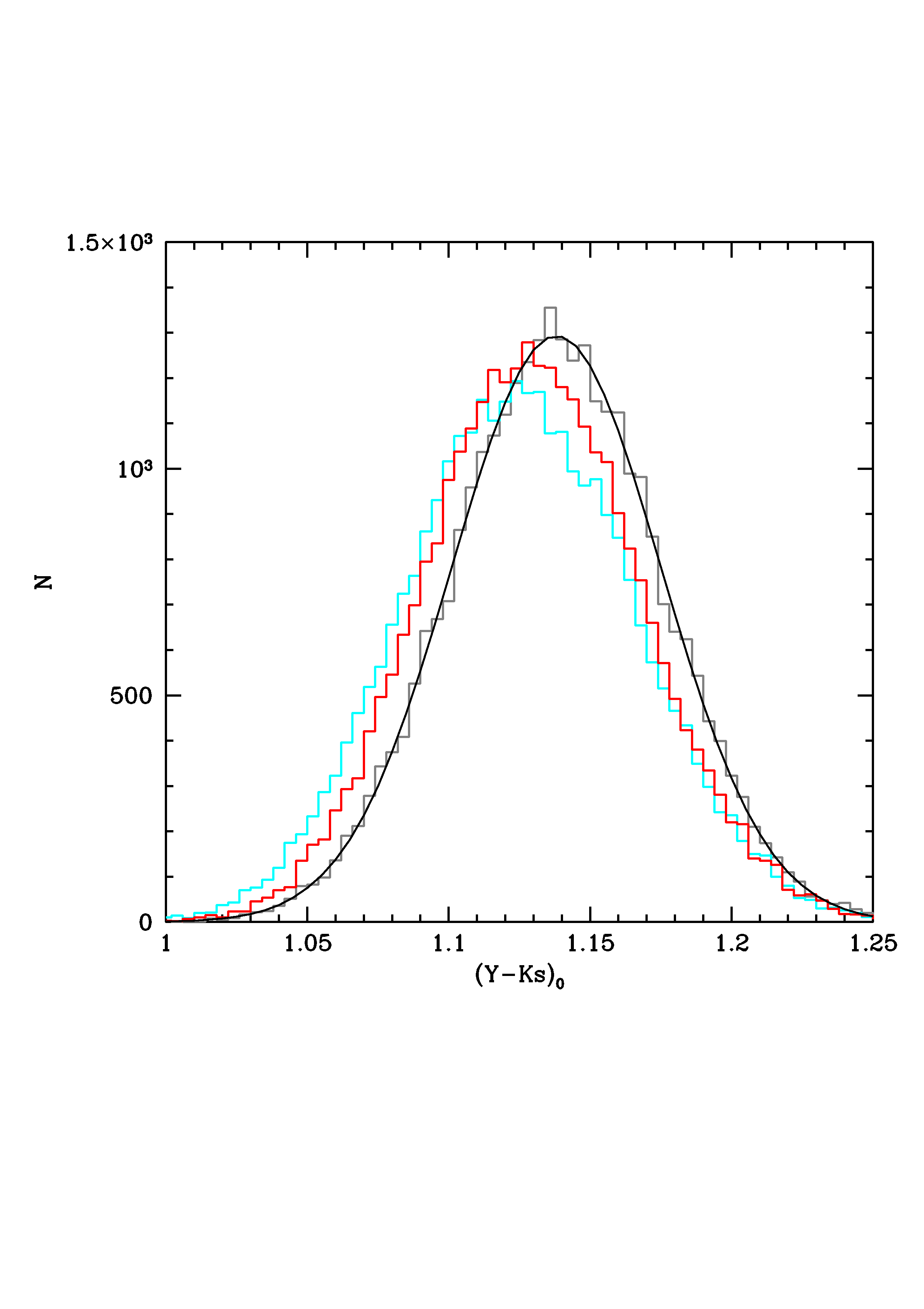}
\caption{The distribution of intrinsic colors $(Z-Ks)_0$ and $(Y-Ks)_0$,
obtained by the statistical bootstrap method for three
strips with a height $|b|<5^\circ$, perpendicular to the Galactic plane with the following longitudes: $l=1.7^\circ$ (black histogram), $l=0.1^\circ$ (red histogram),
$l=2.37^\circ$ (blue histogram). As an example, the black solid line indicates the best-fit Gaussian model for
the strip with $l=1.7^\circ$ (for more details, see the text).}  \label{fig:BS_ZY}
\end{figure*}

Having measured the intrinsic color of RCGs for
the $Z$-filter, we repeated the procedure described
above and determined the true color in the $Y$-filter,
$(Y-K)_0= 1.137\pm0.034$. Given the known $M_{Ks}$,
we obtain the absolute magnitudes of RCGs $M_{Z}=-0.20\pm0.04$ and $M_{Y}=-0.470\pm0.045$, which are
in good agreement with the results of previous sections, but excel them in accuracy.

The proposed method was applied not only for the
strip in the vicinity of Baade’s window, but also for two
more strips with coordinates l = $l=2.37^\circ$ and $l=0.1^\circ$. The
latter passes in the immediate vicinity of the Galactic
center, where the extinction is particularly great and
manifestations of the inconstancy of the extinction
law along the strip, if any, are possible. The results
of our measurements of the intrinsic colors of RCGs
and their absolute magnitudes with corresponding
uncertainties are summarized in Table 2 for all three
strips.

It can be seen from this table that the results
of our measurements for all three strips agree well
between themselves (the corresponding distributions
of dereddened colors $(Z-Ks)_0$ and $(Y-K)_0$ are
shown in Fig. 7).


\begin{table*}

 \centering
\caption{The intrinsic colors and absolute magnitudes of RCGs determined for the sky strips in different parts of the bulge}

\begin{tabular}{c|c|c|c|c}

 \hline
$l$		   & $(Z-Ks)_0$      & $(Y-Ks)_0$      & $M_{Z}$        & $M_{Y}$ \\
 \hline
$1.7^\circ$  & $1.414\pm0.025$ & $1.137\pm0.034$ & $-0.20\pm0.04$ & $-0.470\pm0.045$ \\
$2.37^\circ$ & $1.419\pm0.025$ & $1.130\pm0.032$ & $-0.19\pm0.04$ & $-0.480\pm0.044$	\\
$0.1^\circ$  & $1.420\pm0.027$ & $1.129\pm0.035$ & $-0.19\pm0.04$ & $-0.481\pm0.046$ \\
\hline

\end{tabular}
\end{table*}

\bigskip

To finally demonstrate the operability and properness of the method, we attempted to determine the
absolute magnitudes of RCGs for the filters in which
they are already well known. Since the corresponding magnitudes for the $Ks$ and $J$ filters are a necessary part of our analysis and cannot be derived
based on themselves, we chose the $H$ filter for our
test. As a result of repeating the procedure described
above for the strip in the vicinity of Baade’s window,
we determined the absolute magnitude of RCGs in
this filter, $M_H=-1.50\pm0.03$, that in good agreement with $M_{H}=-1.490\pm0.015$ obtained by Laney et al. (2012) and $M_{H}=-1.49\pm0.04$ measured by Gontcharov (2017).

\section{THE ABSOLUTE MAGNITUDES OF BULGE RED CLUMP GIANTS IN THE
IRAC/SPITZER FILTERS}

Having made sure that the proposed technique is
operable for the filters of the VISTA surveys, we used
it to determine the absolute magnitudes of RCGs in
the filters of the IRAC/Spitzer survey. The study was
performed in the same three sky strips for which the
absolute $Z$ and $Y$ magnitudes were obtained above.
Note that the range of Galactic latitudes covered
by the GLIMPSE/Spitzer surveys is slightly smaller
than the corresponding coverage of the VVV/VISTA
survey, but the number of cells ($170-210$) in each
of the strips turned out to be enough to solve our
problems.
 \begin{figure*}[]
 \centering
\includegraphics[width=1\columnwidth,trim={0.5cm 6cm 0cm 1cm},clip]{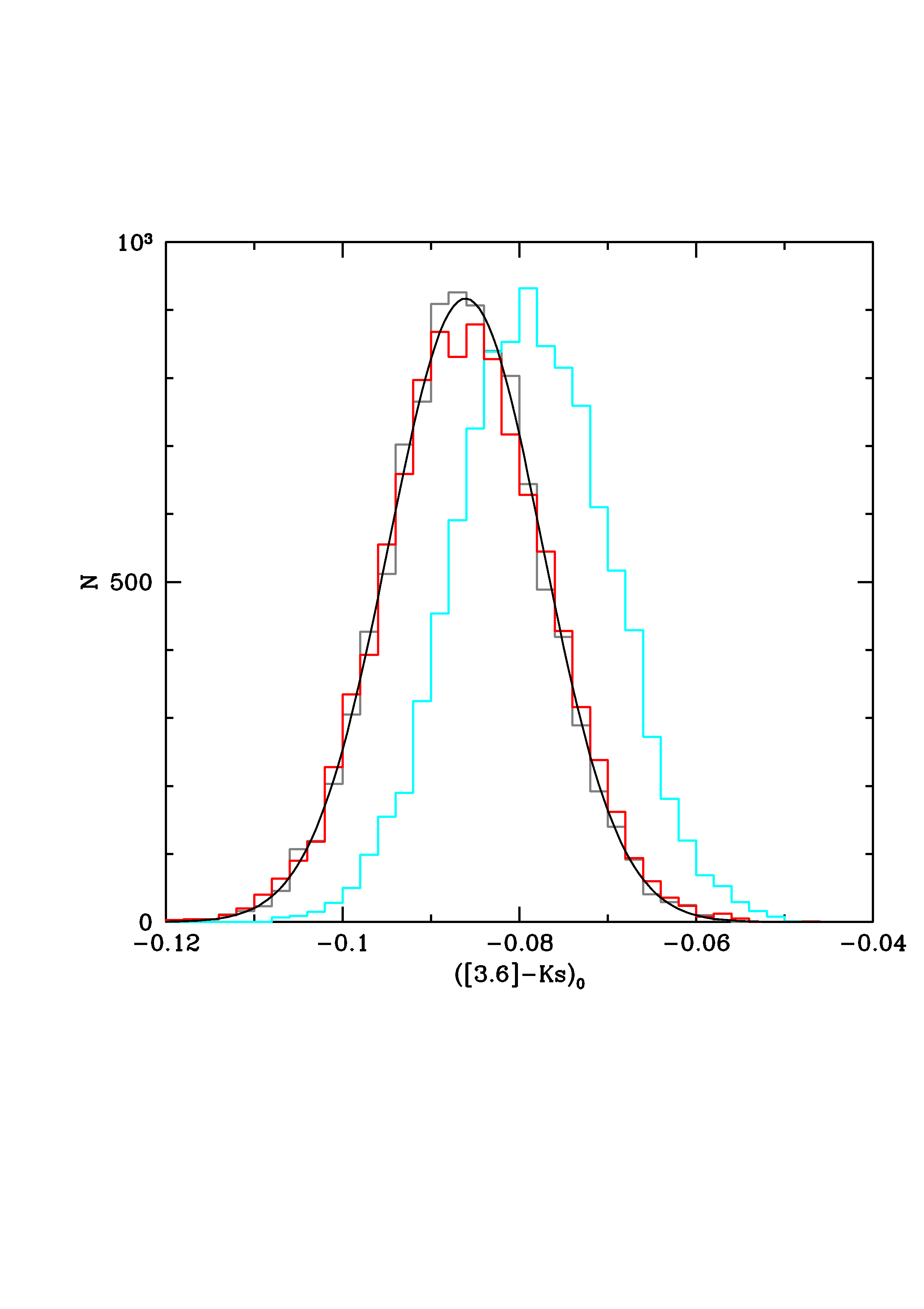}
\includegraphics[width=1\columnwidth,trim={0.5cm 6cm 0cm 1cm},clip]{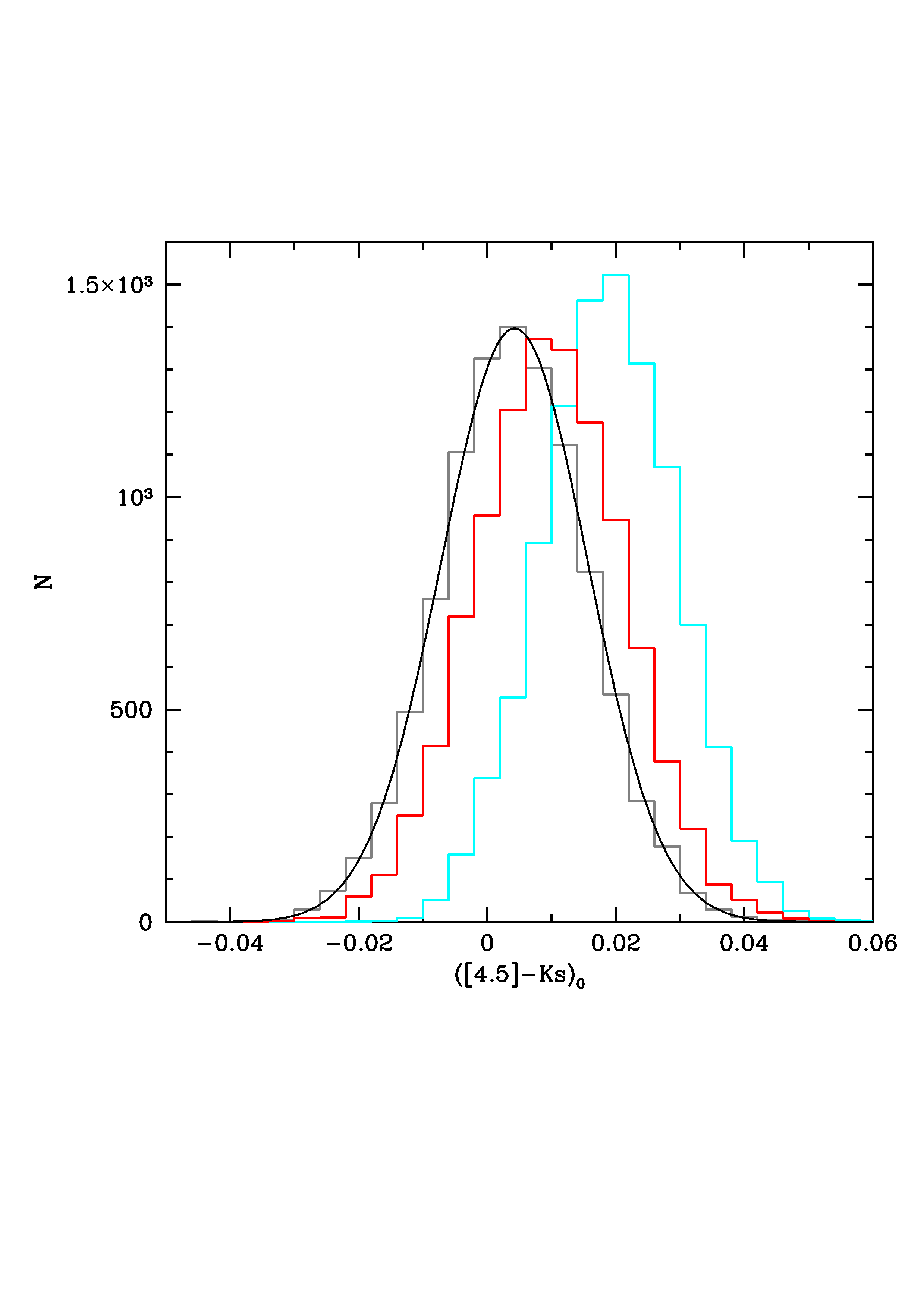}
\includegraphics[width=1\columnwidth,trim={1cm 6cm 0cm 1cm},clip]{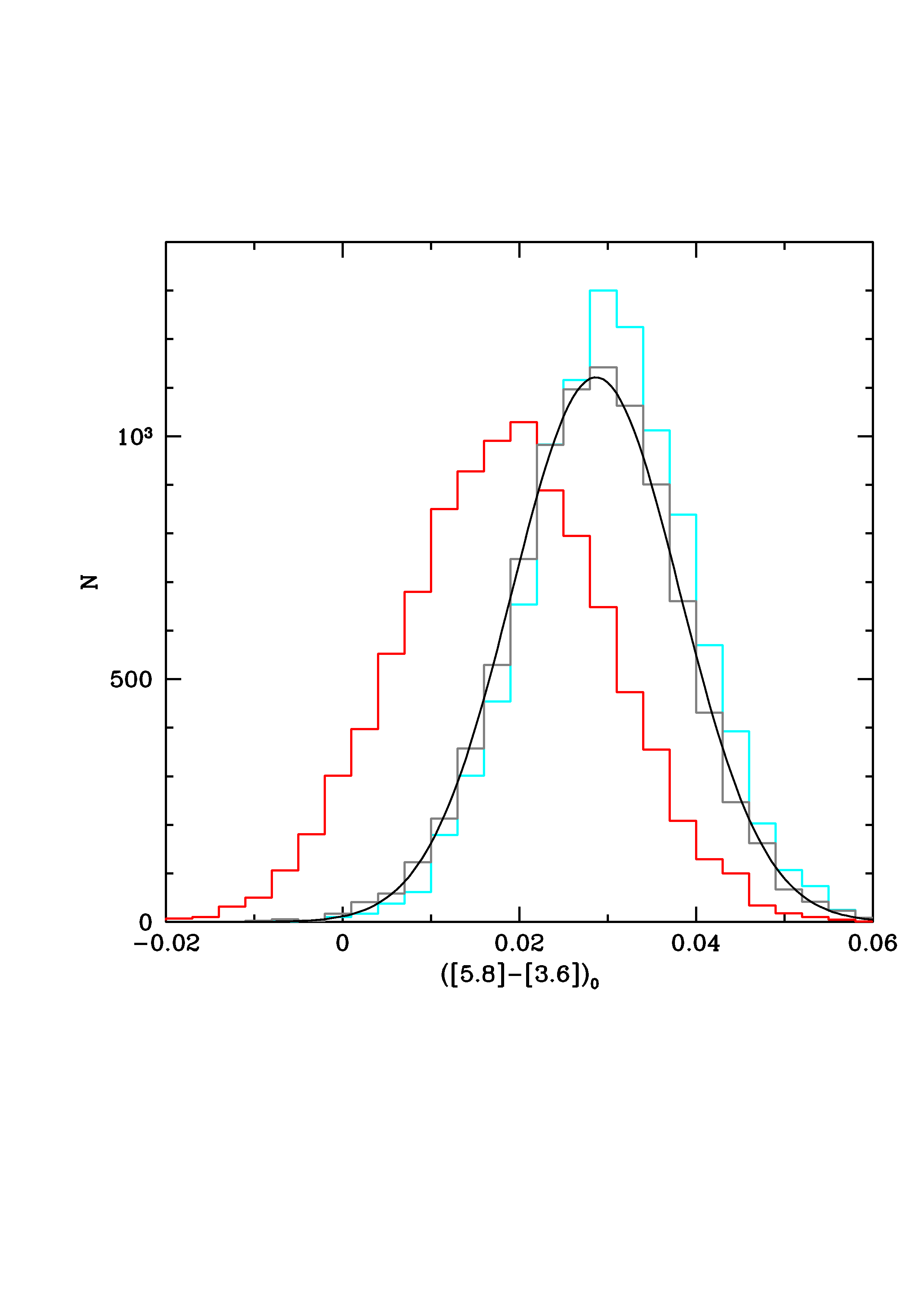}
\includegraphics[width=1\columnwidth,trim={1cm 6cm 0cm 1cm},clip]{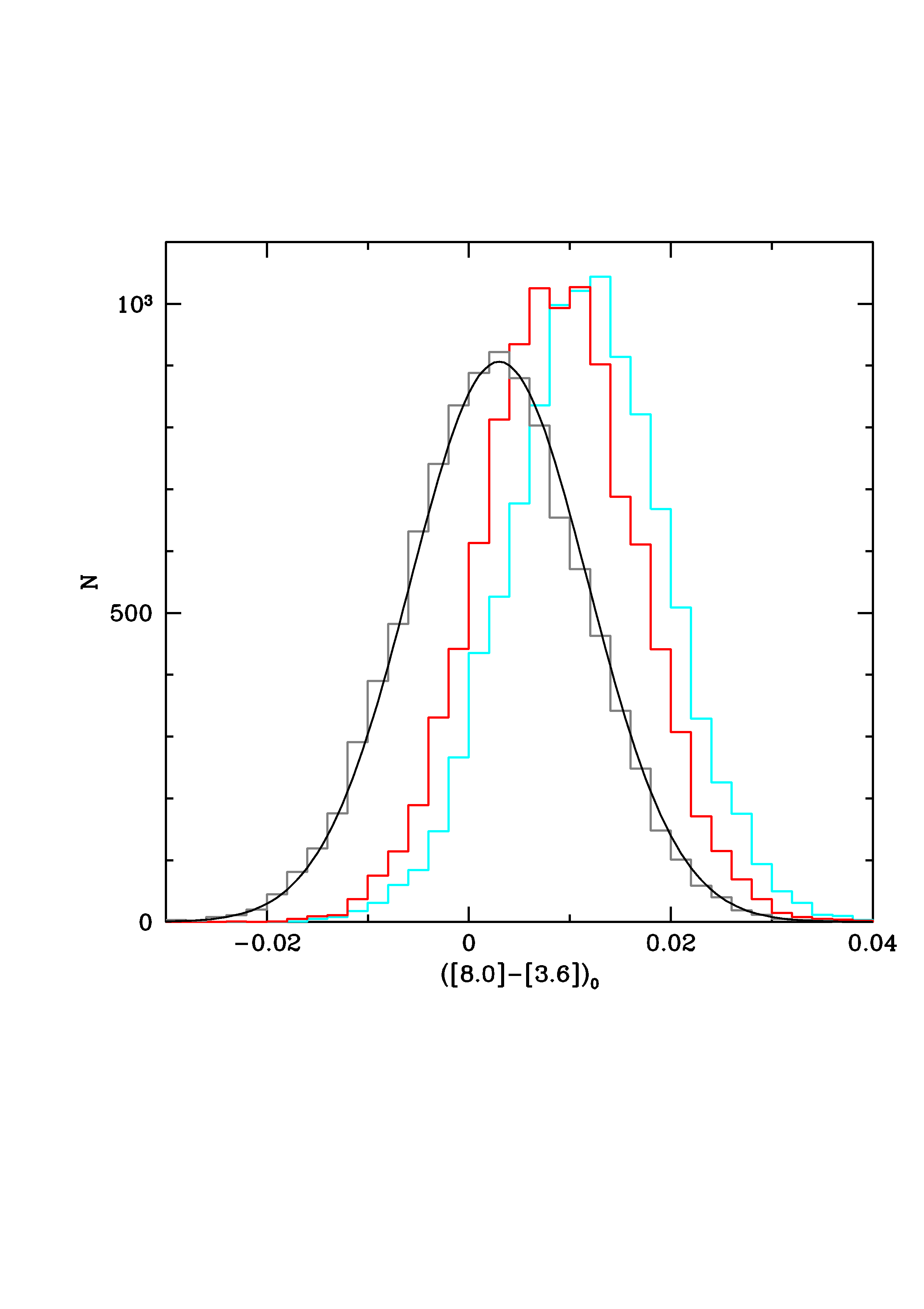}
\caption{Same as Fig. 7 for the GLIMPSE/Spitzer filters.}  \label{fig:BS_SP}
\end{figure*}

To find the absolute magnitudes of RCGs in the
$[3.6]$ and $[4.5]$ filters, as above, we used the known
absolute $Ks$ and $J$ magnitudes and the VVV survey
data in these filters as the basic ones. Therefore, we
first cross-correlated the VVV (DR4) and GLIMPSE
catalogues to select the stars with a statistically significant detection in both catalogues. Then, we determined the colors $([3.6]-Ks)$ and $([4.5]-Ks)$ (the
color $J-Ks$ was determined previously) of the RCG
centroids in the strip cells and used the third method
to determine the intrinsic colors $([3.6]-Ks_0)$ and $([4.5]-Ks)_0$. To find the centroid and to refine the errors, as in the previous section, we used the statistical bootstrap method, except that the number of selected
values was 10 000. The corresponding distributions
are shown in Fig. 8 (upper panels).

As a result of our analysis, we measured the absolute magnitudes of RCGs in the $[3.6]$ and $[4.5]$ filters:
$M_{[3.6]}= -1.70\pm0.03$ and $M_{[4.5]}=-1.60\pm0.03$.

To determine the absolute magnitudes in the $[5.8]$
and $[8.0]$ filters, we used the above results of our measurements of the absolute magnitude in the $[3.6]$ filter.
This is because we managed to determine the colors $([5.8]-[3.6])$ and $([8.0]-[3.6])$ for RCGs more
accurately than the colors $([5.8]-Ks)$ and $([8.0]-Ks)$. Therefore, the third method of determining the
absolute magnitude was applied to these colors and
the color $(J-[3.6])$ as the reference one. The distributions of intrinsic colors $([5.8]-[3.6])_0$ and $([8.0]-[3.6])_0$ are shown in Fig. 8 (lower panels), while the absolute magnitudes of RCGs in the $[5.8]$ and $[8.0]$ filters are $M_{[5.8]}= -1.67\pm0.03$ and $M_{[8.0]}=
-1.70\pm0.03$, respectively. It can be seen from the figure that,
as above, the results for different strips agree well
between themselves, within the uncertainties.

It is also worth noting that in the $[5.8]$ and $[8.0]$
filters of the GLIMPSE survey RCGs are recorded at
the detection limit. Therefore, we admit the presence
of a noticeable admixture of branch giants in our
sample, which can be responsible for some additional
systematic error in the derived magnitudes in these
filters. However, if we look at the isochrones from the
PARSEC database, then we can understand that the
maximum systematic error in the colors used in our
study is small and does not exceed 0.01 mag.

 \begin{figure*}[]
\includegraphics[width=1\columnwidth,trim={1cm 6cm 1cm 1cm},clip]{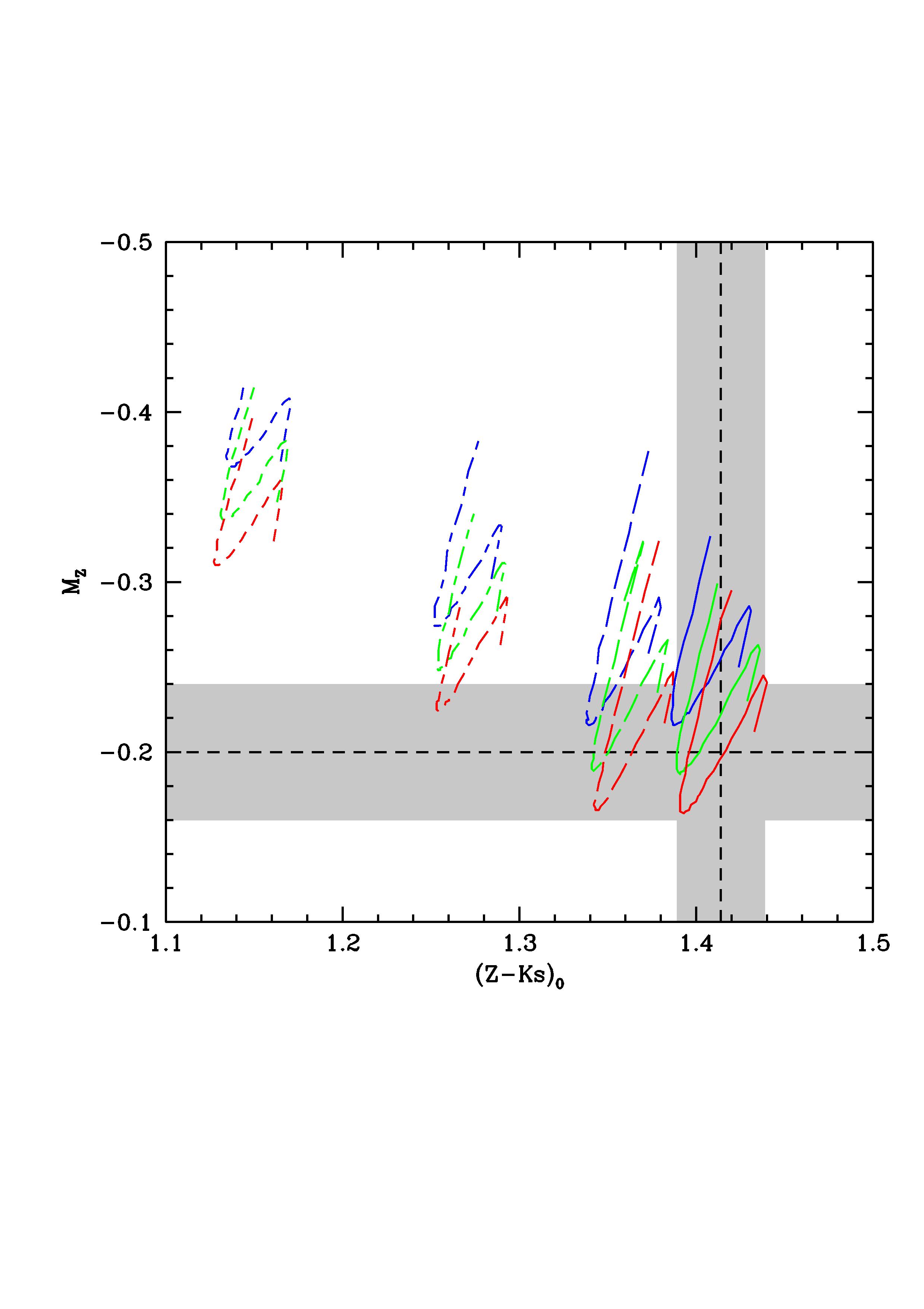}
\includegraphics[width=1\columnwidth,trim={1cm 6cm 1cm 1cm},clip]{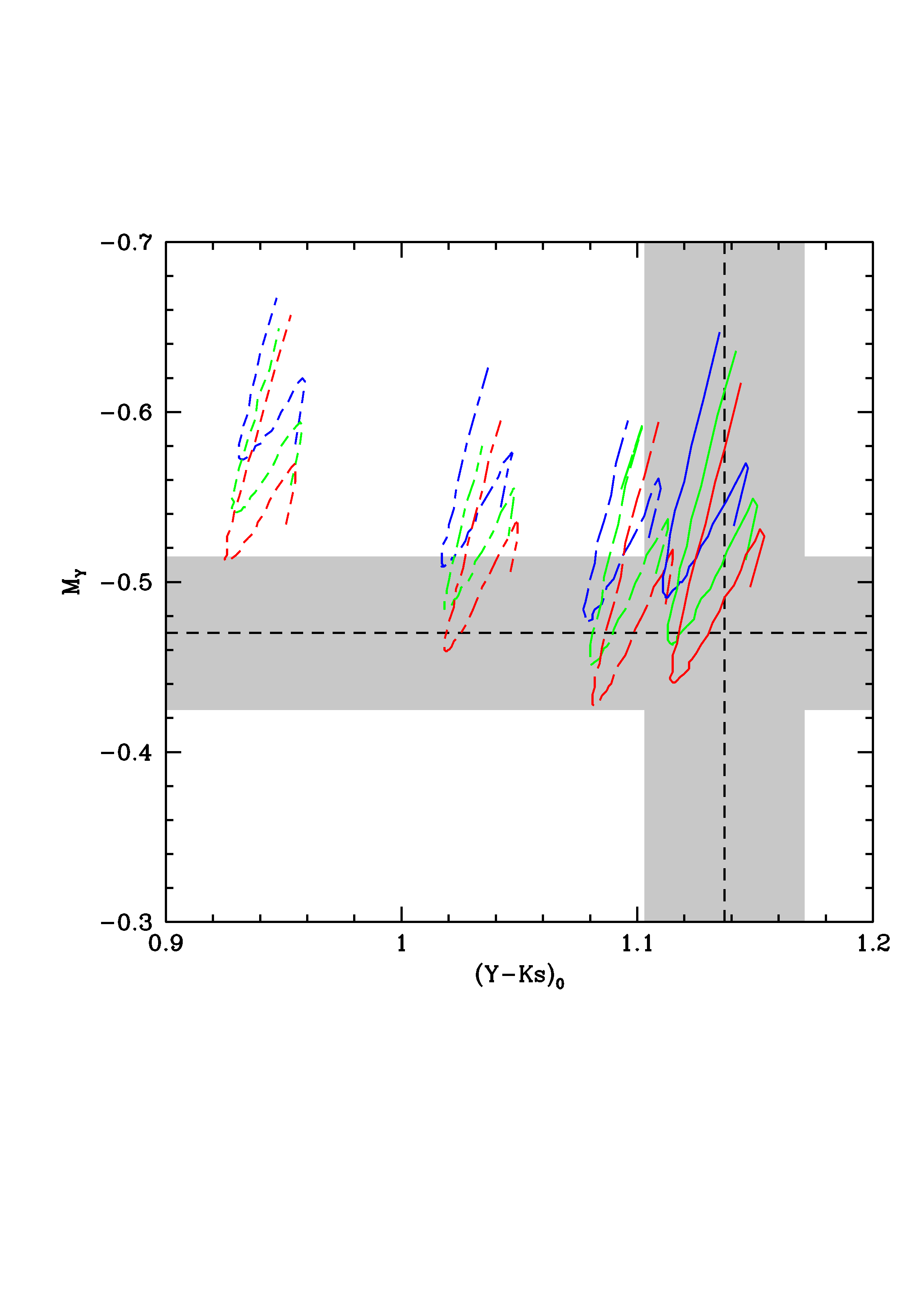}

\caption{ The isochrones for RCGs of different metallicities (${\bf Z}=0.00744,
{\bf Z}_{\sun}=0.0152, {\bf Z}=0.0276, {\bf Z}=0.038$, from left to right) and ages (the red, green, and blue lines for 10, 9, and 8 Gyr, respectively) constructed using the PARSEC
database for the $Z$ filter and color $Z-Ks$ (a) and for the $Y$ filter and color  $Y-Ks$ (b). In the construction we assumed that
there was no interstellar extinction. The black dashed lines mark our absolute values and intrinsic colors. The gray shading
indicates the errors corresponding to one standard deviation for the corresponding magnitudes. The solid lines highlight the
isochrones that correspond best to the observed $Z$ magnitudes.}  \label{fig:PARSEC_ZY}
\end{figure*}

\section{DISCUSSION}

 \begin{figure*}[]
 \centering
\includegraphics[width=1\columnwidth,trim={1cm 6cm 1cm 1cm},clip]{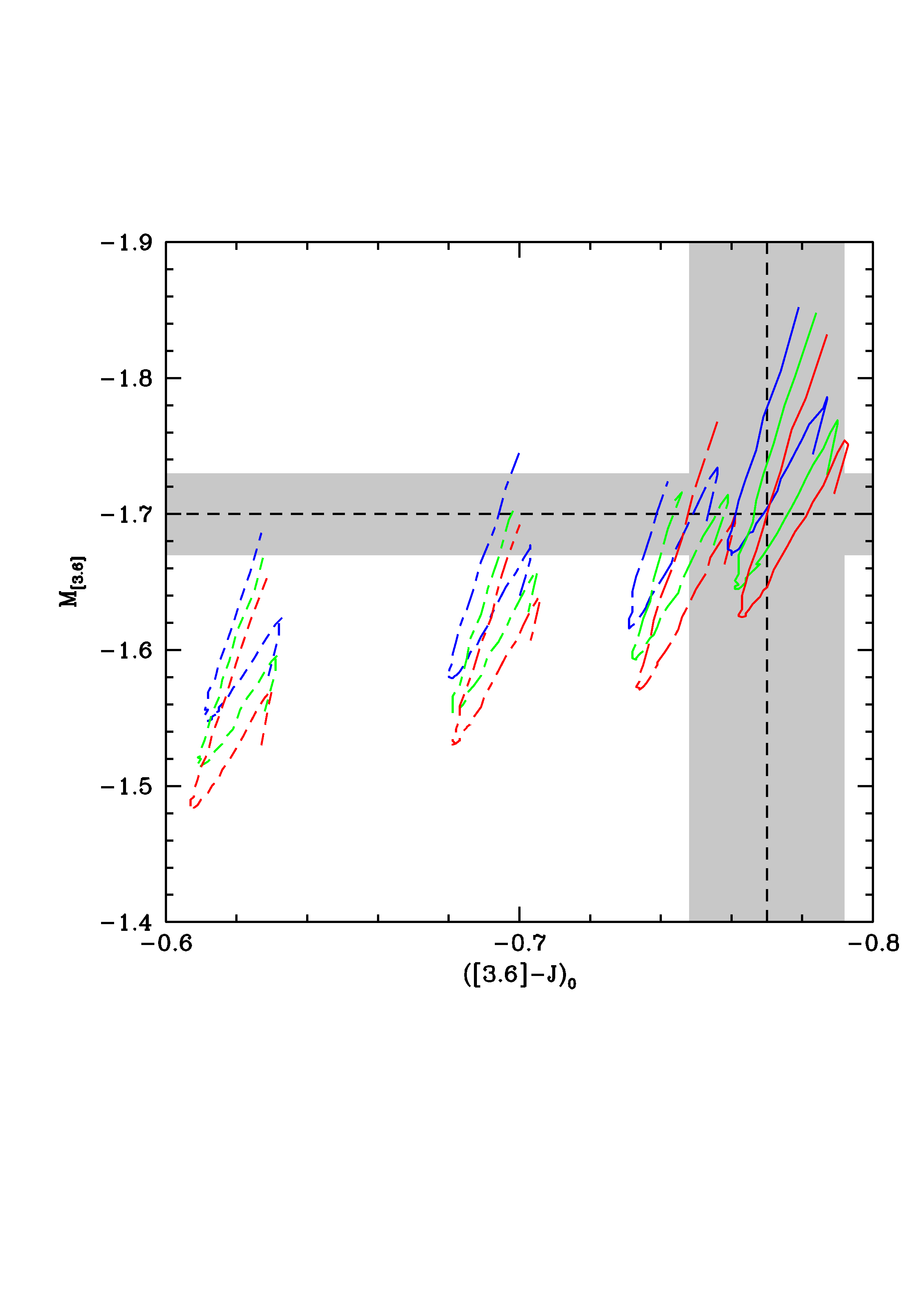}
\includegraphics[width=1\columnwidth,trim={1cm 6cm 1cm 1cm},clip]{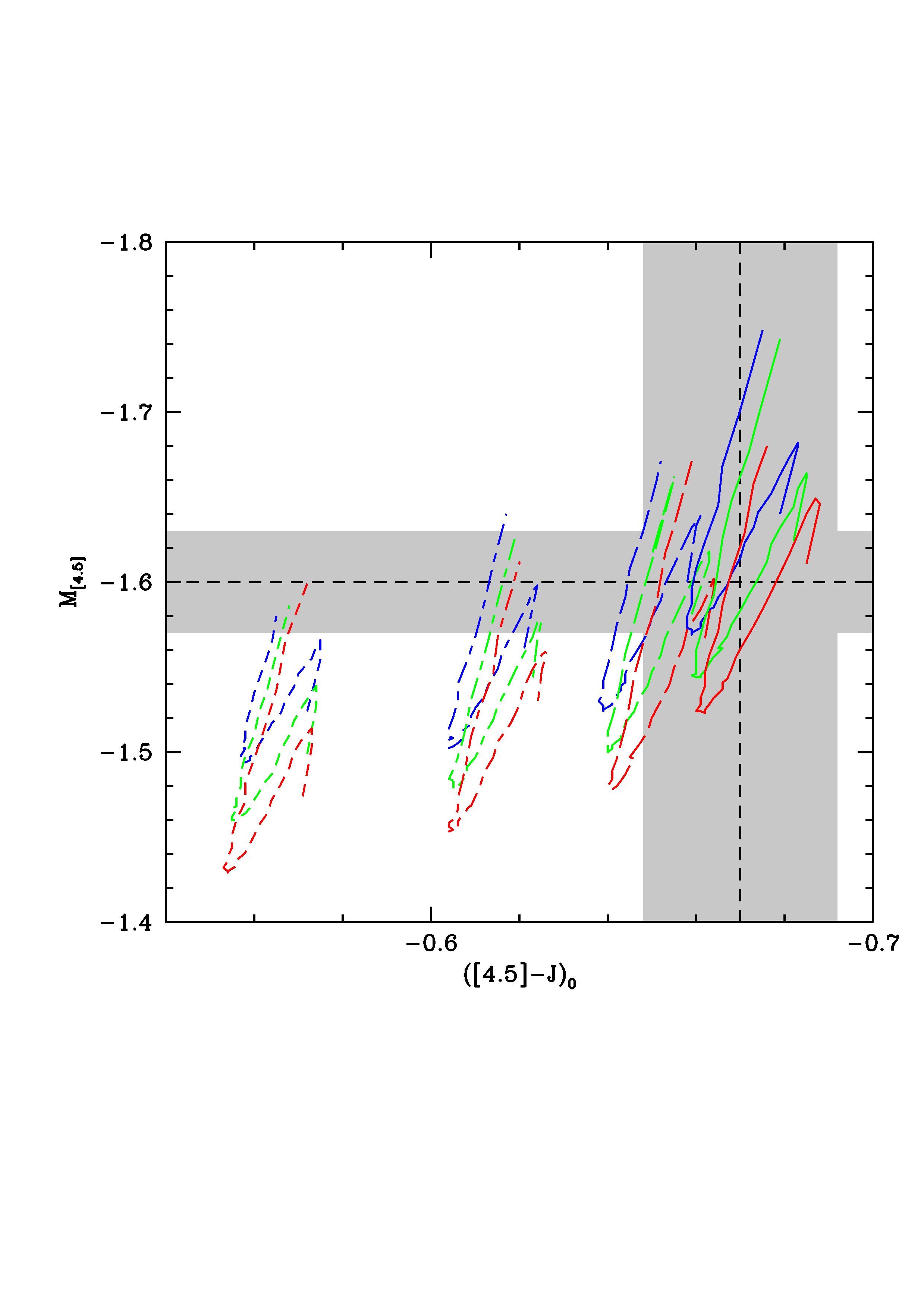}
\includegraphics[width=1\columnwidth,trim={1cm 6cm 1cm 1cm},clip]{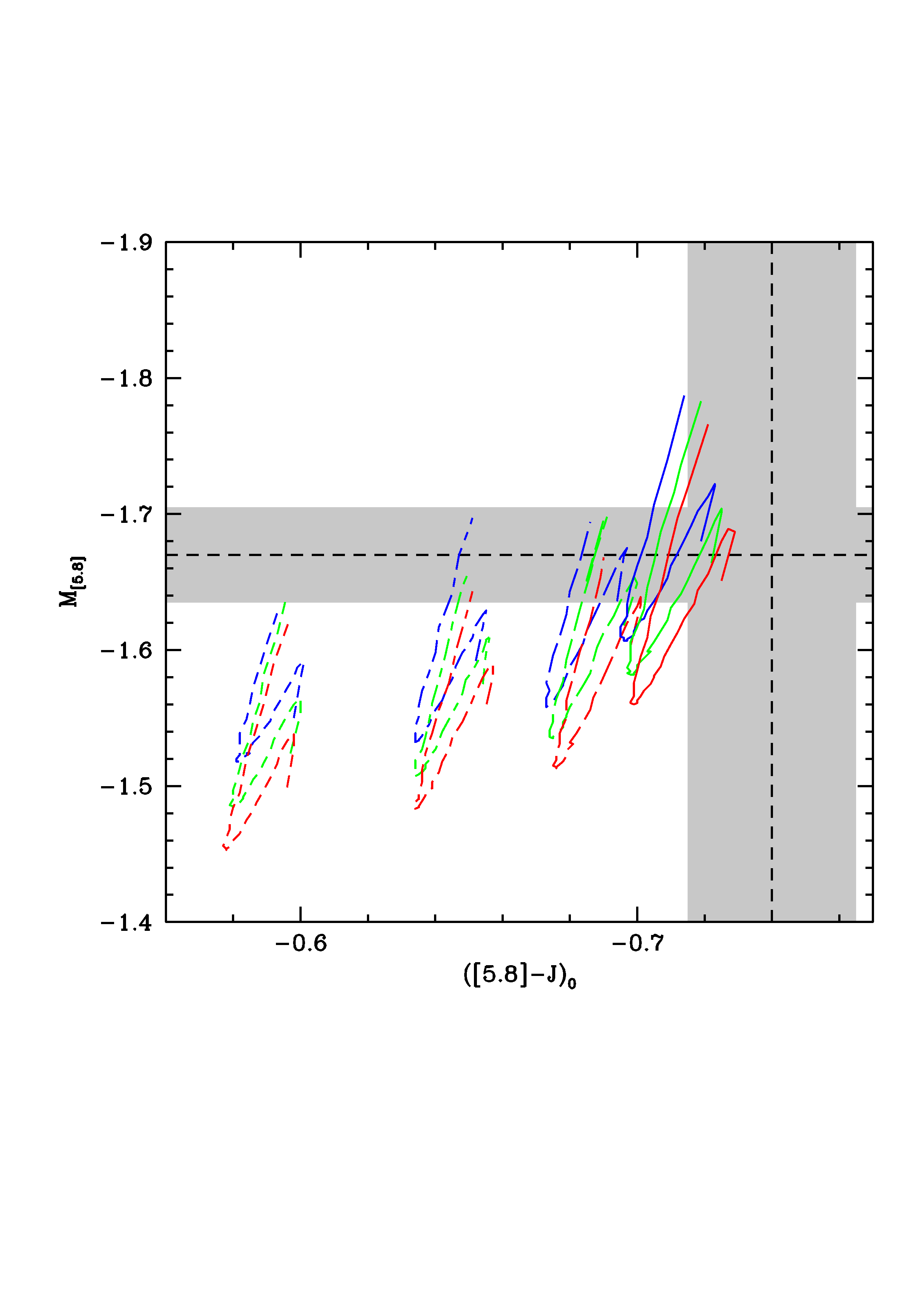}
\includegraphics[width=1\columnwidth,trim={1cm 6cm 1cm 1cm},clip]{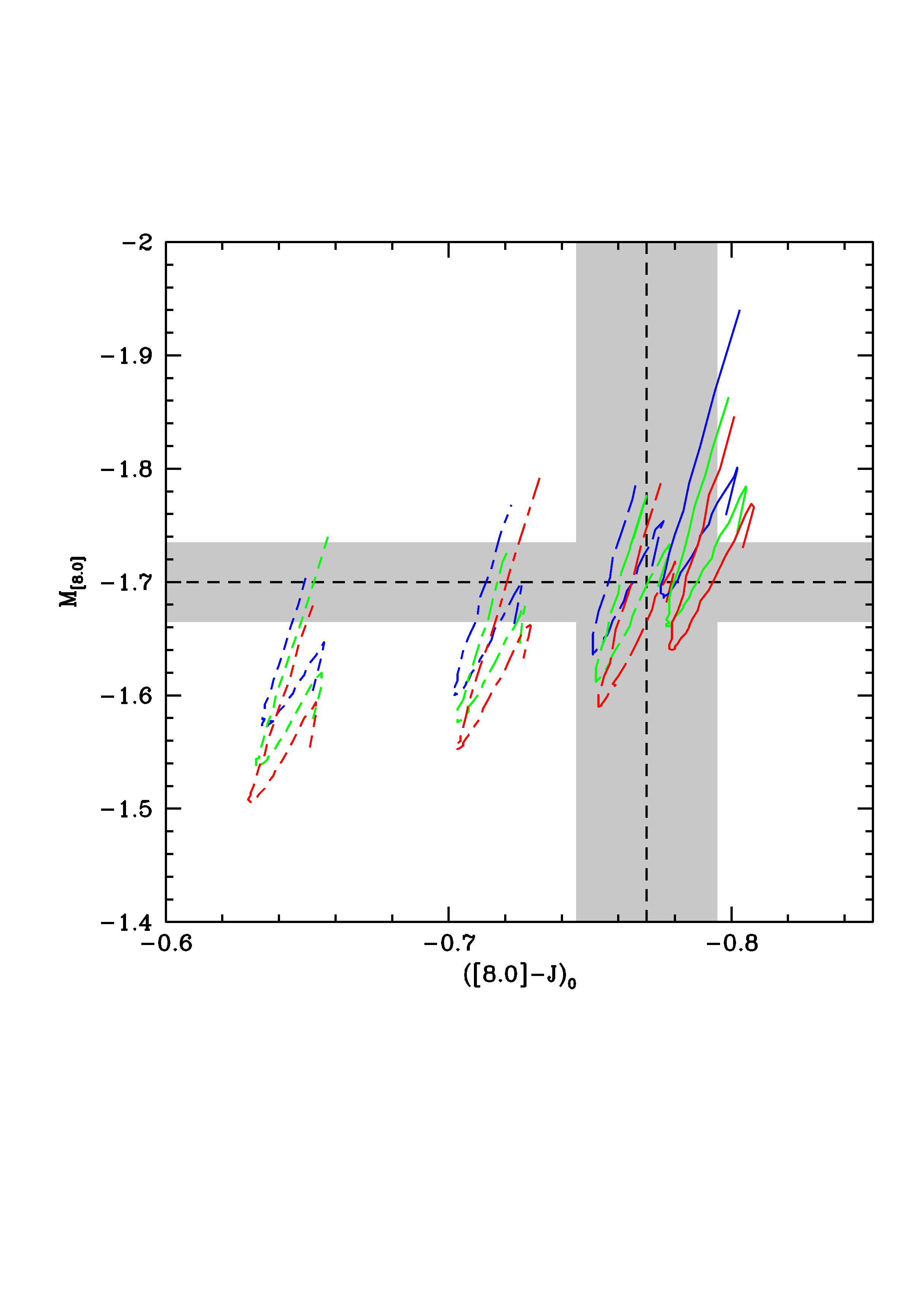}
\caption{Same as Fig. 9 for the GLIMPSE/Spitzer filters.}  \label{fig:PARSEC_SP}
\end{figure*}
To understand how well our absolute magnitudes
of RCGs agree with theoretical models, we compared
them with the PARSEC isochrones constructed for
the ages and metallicities typical of bulge RCGs. We
chose the isochrones for RCGs of three ages, 8, 9, and
10 Gyr (see, e.g., Zoccali et al. 2003; Vanhollebeke
et al. 2009), and two metallicities from Gonzalez
et al. (2015) -- $[Fe/H]\simeq-0.40$ (the low-metallicity
subgroup of RCGs) and $[Fe/H]\simeq0.26$ (the high-metallicity
subgroup of RCGs), as well as the solar metallicity
(abundance). Since a different parametrization of
the metallicity (${\bf Z}$) is used in the PARSEC database
and the isochrones normalized to the solar abundance
are presented, below we adopt $[M/H]=[Fe/H]$. In
this case, $[M/H]$ were recalculated to {\bf Z} using the
formula $[M/H]=log_{10}({\bf Z}/{\bf Z_{\sun}})$ by assuming ${\bf Z}_{\sun}=0.0152$, according to the recommendations of the
PARSEC database. Thus, the above metallicities
from Gonzalez et al. (2015) correspond to ${\bf Z}=0.00744$ and ${\bf Z}=0.0276$. The results of the comparison
of the isochrones and our measurements are shown
in Figs. 9 and 10.

It can be seen that our absolute magnitudes in
all filters agree with the theoretical models for high-
metallicity RCGs. Moreover, if add the intrinsic
colors of RCGs in different filters to our analysis,
then additional constraints on the metallicity of bulge
RCGs can be obtained. In particular, to correspond
to the measured values within one standard deviation,
their metallicity must be $[M/H]\simeq 0.4$ (${\bf Z} = 0.038$)
with an uncertainty no greater than $0.1 dex$. Although
this is formally slightly larger than that in Gonzalez
et al. (2015), it agrees with the estimates of the latter
within the error limits.

It can also be seen from the comparison of the
derived magnitudes with the isochrones that objects
with an age of 9--10 Gyr, with the possible inclusion
of younger objects with an age of $\sim$8 Gyr, must
dominate among the bulge RCGs. The $Z$ filter close
to the optical band, which allows the most accurate
estimates to be obtained, is most revealing in this
regard. The measurements in the $Y$ , [3.5], and [4.5]
filters also agree well with the above estimates; the results for the more distant [5.6] and [8.0] filters give less
accurate agreement, but they are consistent within
the measurement uncertainty limits.

 \begin{figure}[]
 \begin{centering}
\includegraphics[width=1\columnwidth,trim={1cm 6cm 1cm 1cm},clip]{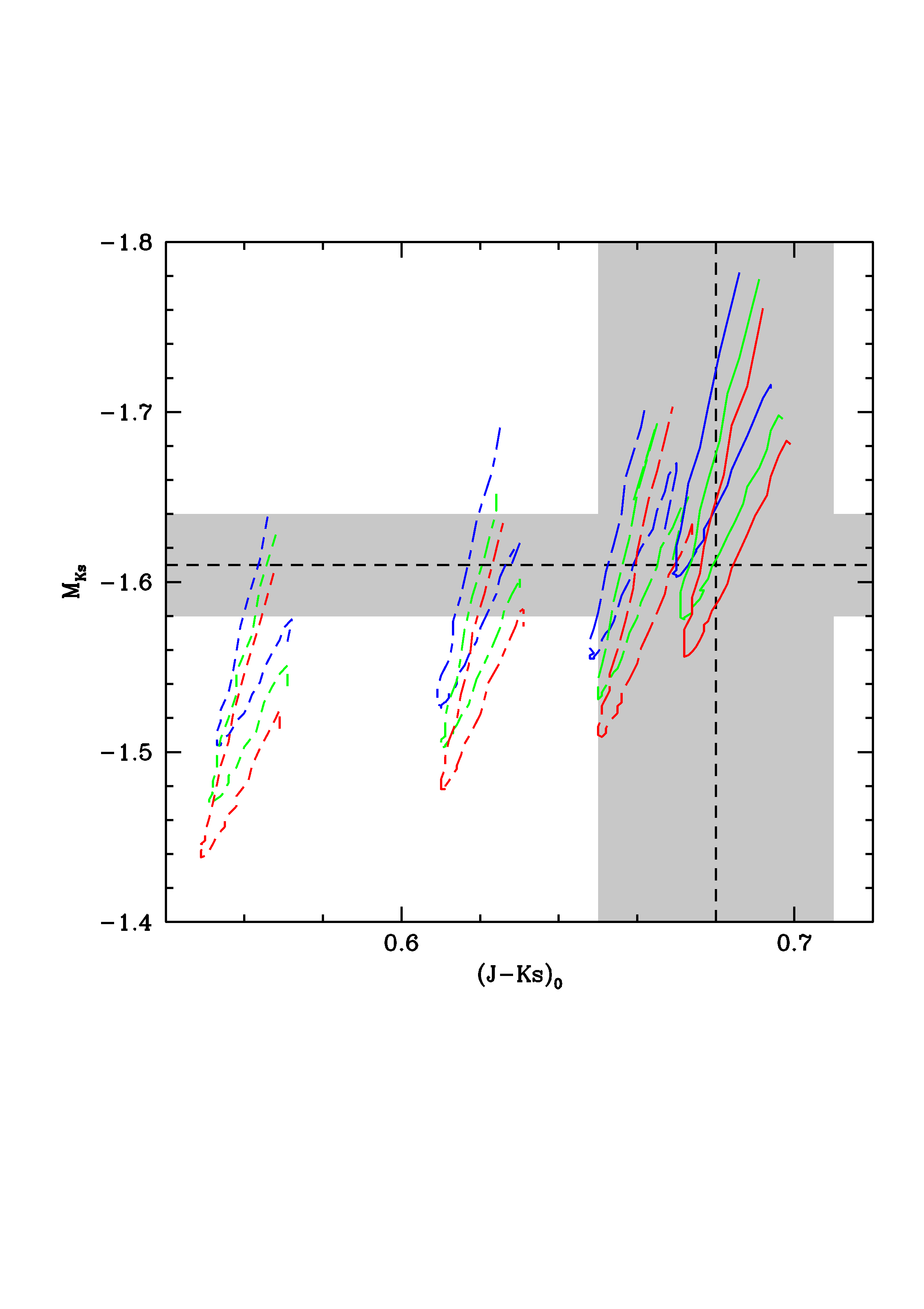}
\caption{Same as Fig. 9 for the J and Ks filters.}
\label{fig:PARSEC_JK}
\end{centering}
\end{figure}

Note that the natural scatter of RCG colors following from the PARSEC isochrones is comparable to our standard deviations for the colors of these
objects, i.e., our estimates correspond in accuracy to
the "natural" accuracy (the scatter of colors) typical for RCGs.

It is interesting to check how well the absolute
magnitudes of RCGs in the J and Ks filters, which
we everywhere used as the reference ones, agree with
our metallicity estimates. Using the PARSEC data,
we constructed the isochrones for RCGs of three
ages and four metallicities (as above) in these filters
(Fig. 11). It follows from the figure that the reference
values for RCGs in the $J$ and $Ks$ filters agree excellently with the theoretical expectations for our metallicity. The metallicity $[Fe/H]\simeq0.26$ from Gonzalez
et al. (2015) is also consistent with our results and
falls into the range of measurement errors.

Using the estimated absolute magnitudes of RGCs
and their metallicities, we checked how well our
results agreed with the stellar atmosphere models
for these objects. For this purpose, we converted
the derived magnitudes to the flux densities and
compared them with the spectra of red giants with
temperatures from $\sim$4000 to $\sim$4750 K (this choice is
explained by the universally accepted view of RCG
temperatures; see, e.g., Girardi 2016) generated with
the SYNPHOT package for synthetic photometry
and the Kurucz library of spectra for ${\bf Z} = 0.038$ and ${\bf Z} = 0.0152$ (solar abundance).
The result of such a comparison is shown in Fig. 12. We used the relations
from Cohen et al. (2003) to convert the $J$, $H$, $Ks$
magnitudes and the relations at the official VISTA
site\footnote{\url{http://casu.ast.cam.ac.uk/surveys-projects/vista/technical/filter-set}} (they were first reduced to the AB system using
the corresponding coefficients) to convert the $Z$ and
$Y$ magnitudes. To convert the Spitzer magnitudes,
we used the resource\footnote{\url{http://svo2.cab.inta-csic.es/svo/theory/fps/index.php?mode=browse&gname=Spitzer}} and conversion tools at the
Gemini site\footnote{\url{http://www.gemini.edu/sciops/instruments/midir-resources/imaging-calibrations/fluxmagnitude-conversion}}.

It can be seen that our results agree well with the
assumptions about the temperature properties of red
giants. Moreover, using the estimates for the metallicity of these stars obtained above, we can determine the most suitable temperature for the RCGs under
study, $4250\pm150$\,K (Fig. 12).

\begin{figure*}[]
\begin{centering}
\includegraphics[width=1.5\columnwidth,trim={1cm 6cm 0.5cm 0.5cm},clip]{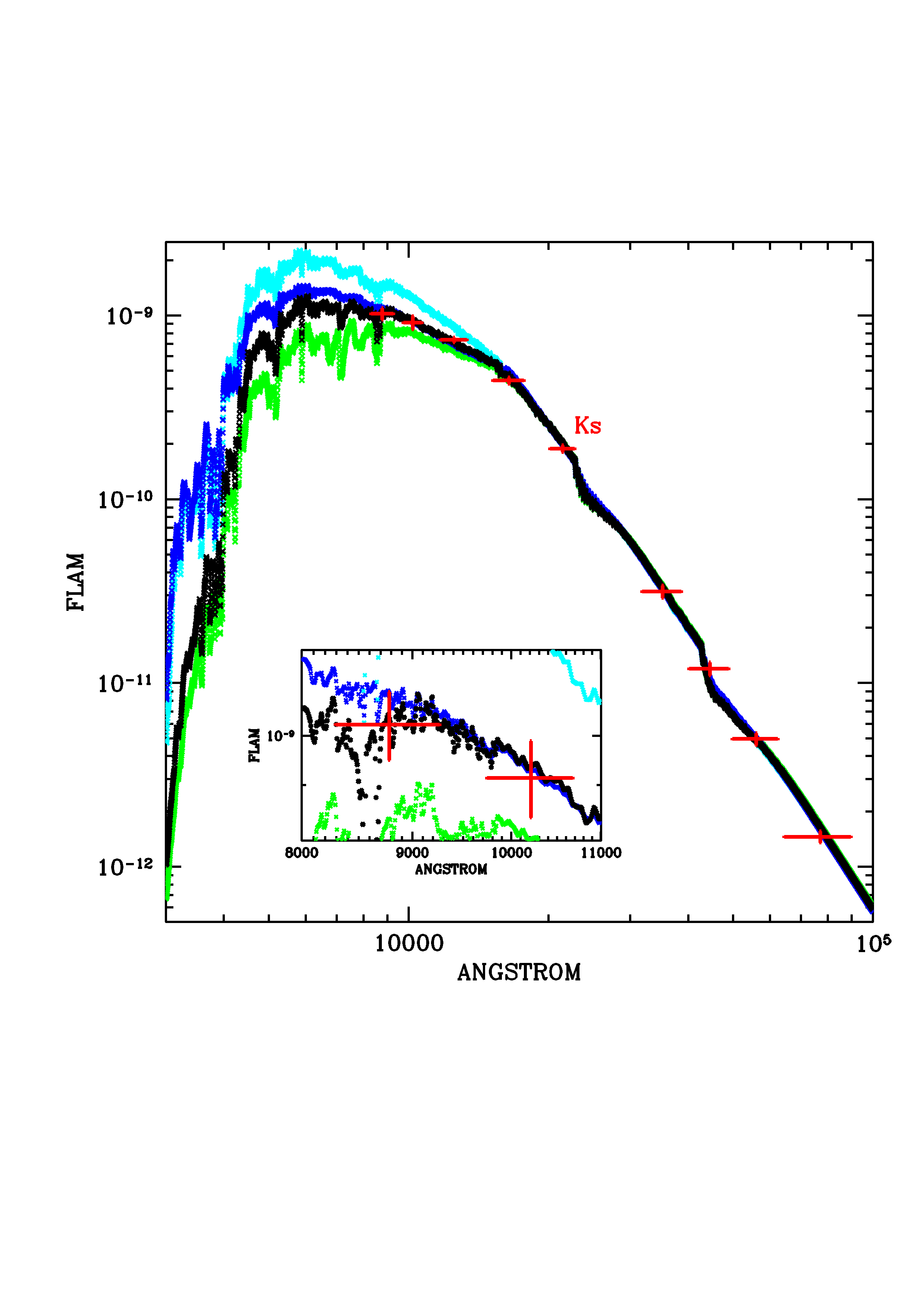}
\caption{The synthetic spectra constructed for red giants with temperatures of $4000$\,K (green curve), $4250$\,K (black curve), $4750$\,K (blue curve) for metallicity ${\bf Z} = 0.038$, and $4400$\,K for the solar abundance ${\bf Z} = 0.0152$ (blue curve) using the {\sc SYNPHOT} package and the \emph{Kurucz} library of spectra. The spectra were normalized to the absolute $Ks$ magnitude of RCGs. The crosses indicate the fluxes corresponding to the absolute magnitudes of RCGs in nine infrared filters: $Z, Y, J, H, Ks, [3.6], [4.5], [5.8] and [8.0]$ (from left to right), including those measured in this paper. A magnified part of the main picture near the $Z$ and $Y$ filters is presented in a separate inset.}
\label{fig:ZY_SPEC}
\end{centering}
\end{figure*}

\section{CONCLUSIONS}

In this paper for the first time we have determined the absolute magnitudes of bulge red clump giants in the infrared $Z$ and $Y$ filters of the VVV
(VISTA/ESO) survey and the [3.6], [4.5], [5.8], and
[8.0] bands of the GLIMPSE (IRAC/Spitzer)
Galactic-plane survey. Our results are based on the
use of narrow vertical strips to determine the colors of
RCGs in different filters and the extinction law. This
method was proposed by Karasev et al. (2015) and
developed further in this paper.

The results obtained in this paper (the absolute
magnitudes, the corresponding flux densities, and
their standard deviations) are presented in Table 3.

A comparison of the derived absolute magnitudes
with theoretical models allowed us to measure the
metallicity of bulge RCGs, $[M/H]\sim0.40$ (or ${\bf Z} \sim 0.038$) with an error of $[M/H]\sim 0.1$ dex, and their characteristic temperature, $\sim
4250\pm150$ K. We also
showed that stars with an age of 9 -- 10 Gyr dominate
among RCGs, but, at the same time, younger objects
with an age of $\sim$8 Gyr may also be present.

 \begin{table*}
 \centering
  \footnotesize{
  \caption{Parameters of the photometric bands under study and the absolute magnitudes of bulge RCGs}

      \begin{tabular}{c|c|c|c|c}

  \hline
 Filter 	&   $\lambda_{eff}, $ \AA  &   $\omega,$ \AA   &    $M$, mag   & FLAM, erg cm$^{-2}$ s $^{-1}$ \AA$^{-1}$ \\

 \hline

$Z_{Vista}$  & 8762.58   &   970  &  $-0.20\pm0.04$  &	$   1.024_{\pm0.07}\times 10^{-09} $\\
$Y_{Vista}$  & 10239.64   &  930   &    $-0.470\pm0.045$  &  $	9.14_{\pm0.74}\times 10^{-10} $ \\
$[3.6]_{Spitzer}$  & 35242.46   &   6836  &    $ -1.70\pm0.03$  & $ 3.15_{\pm0.24}\times 10^{-11}  $\\
$[4.5]_{Spitzer}$  & 44540.45   &   8649  &    $-1.60\pm0.03$  & $1.20_{\pm0.11}\times 10^{-11} $\\

$[5.8]_{Spitzer}$ & 56095.61   &   12561.17  &    $-1.67\pm0.03$  & $4.96_{\pm0.37}\times 10^{-12} $\\
$[8.0]_{Spitzer}$ & 77024.24   &   25288.50  &    $-1.70\pm0.03$  & $1.46_{\pm0.1}\times 10^{-12} $\\

\hline
$H$  & 16372.50   &   2510  &    $-1.50\pm0.03$  &$  7.37_{\pm0.22}\times 10^{-10}  $\\

\hline

\hline

\end{tabular}
}
\end{table*}

It is important to note the absence of noticeable
differences in the absolute magnitudes derived for
strips in different parts of the Galactic bulge. This
may suggest that the metallicity in the vicinity of the
Galactic center differs only slightly from the metallicity near Baade's window. In addition, good agreement
between the absolute magnitudes of RCGs (primarily
in the $Y$ filter) obtained by different methods, including those assuming the constancy of the distance and
extinction law along the entire strip, is worth noting.
This may suggest that the extinction law is highly
constant within the chosen strips perpendicular to the
Galactic plane. On the other hand, the infrared filters
may simply less sensitive to small variations in this
law (Nataf et al. 2013, 2016).

We also showed that toward the sky regions under
study the extinction law differs noticeably from the
standard one, and its allowance is very important for
obtaining the correct estimates of the absolute magnitudes. In addition, we estimated the distances to
several Galactic bulge regions, -- $D=8200 - 8500$ pc,
in good agreement with the results of other papers.

\vspace{15mm}
  \section*{ACKNOWLEDGMENTS}
We are grateful to G. Gontcharov for a number of valuable remarks and constructive suggestions that helped improve the paper significantly. We also thank the anonymous referee for useful re- marks. We used the VVV data obtained with the VISTA/ESO telescope (Paranal Observatory) within program ID 179.B-2002 and data from the Spitzer Space Telescope maintained by the Jet Propulsion Observatory of the California Institute of Technology and NASA. This work was financially supported by the Program of the President of the Russian Federation for support of leading scientific schools (project no. NSh-10222.2016.2) and the “Origin, Structure, and Evolution of Objects in the Universe” Program of the Presidium of the Russian Academy of Sciences. D.I. Karasev also thanks the Russian Foundation for Basic Research (project no. 16-02-00294), while A.A. Lutovinov thanks the “Dynasty” Foundation for its partial support of this work.

\section*{REFERENCES}
\noindent

1. D.R. Alves, \apj  {\bf 539}, 732 (2000).\\

2. J. Alonso-Garcia, D. Minniti, M. Catelan, R. Ramos, O. Gonzalez, M. Hempel, P. Lukas,  R. Saito, et al,  arXiv:1710.04854v1 (2017). \\

3. T.-L. Astraatmadja and C.A.L. Bailer-Jones , \apj  {\bf 833}, 119 (2016). \\

4. Boldin P.A., Tsygankov S.S., Lutovinov A.A.,  Astron. Lett.  {\bf 39}, 375 (2013). \\

5. T. Bensby, J.C. Yee, S. Feltzing, J.A. Johnson, A. Gould, J.G.Cohen, M. Asplund, J. Melendez, et al., \aap {\bf 549}, 147 (2013). \\

6. A. Bressan, P. Marigo, L. Girardi, B. Salasnich, C. Dal Cero, S. Rubele, A. Nanni, MNRAS {\bf 427}, 127 (2012), http://stev.oapd.inaf.it/cmd. \\

7. A. Bhardwaj,  M. Rejkuba, D. Minniti,  F. Surot, E. Valenti, M. Zoccali, O. A. Gonzalez, M. Romaniello, et al, \aap   {\bf 605}, id.A100 (2017). \\

8. E. Vanhollebeke, M. A. T Groenewegen, L. Girardi, \aap  {\bf 498}, 95 (2009). \\

9. O. Gerhard,  Martinez-Valpuesta,  Astrophys. J. Letters {\bf 744}, L8,  (2012). \\

10. O. A. Gonzalez, M. Rejkuba,  M. Zoccali, E. Valenti, D. Minniti, M. Schultheis, R. Tobar, B. Chen, \aap  {\bf 552}, 9 (2012). \\

11. O. A. Gonzalez, M. Zoccali, S. Vasquez, V. Hill, M. Rejkuba, E. Valenti, \aap {\bf 584}, A46, (2015). \\

12. Gontcharov G.A., Astron. Lett. {\bf 34}, 785 (2008). \\

13. Gontcharov G.A., Astron. Lett. {\bf 38}, 12 (2012). \\

14. Gontcharov G.A. and  Baykova A.T., Astron. Lett. {\bf 39}, 689 (2013). \\

15. Gontcharov G.A., Astron. Lett. {\bf 43}, 545, (2017). \\

16. C.M. Dutra,   B.X. Santiago, E.L.D. Bica, B. Barbuy, MNRAS {\bf 338}, 253 (2003). \\

17. Leo Girardi, MNRAS {\bf 308}, 818-832 (1999). \\

18. Leo Girardi, Ann. Rev. Astron.  Astrophys. {\bf 54}, p.95-133 (2016). \\

19. M. Zoccali, A. Renzini, S. Ortolani, L. Greggio, I. Saviane, S. Cassisi, M. Rejkuba, B. Barbuy, R. M. Rich and E. Bica, \aap {\bf 399}, 931 (2003). \\

20. Karasev, D. I., Lutovinov, A. A., Burenin , R. A.,  MNRAS Lett. {\bf 409}, L69 (2010a). \\

21. Karasev, D. I., Revnivtsev, M. G., Lutovinov, A. A., Burenin, R. A.,  Astron. Lett. {\bf 36}, 788 (2010b). \\

22. Karasev, D. I., Tsygankov, S. S., Lutovinov, A. A., Astron. Lett. {\bf 41}, 394 (2015). \\

23. J.A. Cardelli, G.C. Clayton  and J.S. Mathis, \apj {\bf 345}, 245 (1989). \\

24. M. Cohen, Wm. A. Wheaton, S. T. Megeath, \apj {\bf 126}, 1090 (2003). \\

25. C. D. Laney, M. D. Joner,  G. Pietrzy'nski, MNRAS {\bf 419}, Issue 2 (2012). \\

26. P. Marigo, L. Girardi, A. Bressan, Philip Rosenfield, Bernhard Aringer, Yang Chen, Marco Dussin, Ambra Nanni, et al, \apj {\bf 835}, 77  (2017). \\

27. D. M. Nataf,  A. Gould,  P. Fouque', O. A. Gonzalez, J. A. Johnson, J. Skowron, A. Udalski, M. K. Szymanski, et al , \apj {\bf 769}, 88 (2013). \\

28. D. M. Nataf,  O. A. Gonzalez,  L. Casagrande, G. Zasowski, C. Wegg, C. Wolf, A. Kunder, J. Alonso-Garcia, et al., MNRAS {\bf 456}, 2692 (2016). \\

29. S. Nishiyama,  M. Tamura,  H. Hatano, D. Kato, T. Tanabe, K. Sugitani, T. Nagata, \apj {\bf 696}, 1407 (2009). \\

30. B. Paczynski and K. Stanek, \aap {\bf 494}, 219 (1998). \\

31. P. Popowski, \apj {\bf 528}, 9 (2000). \\

32. Revnivtsev, M., van den Berg, M., Burenin, R., Grindlay, J. E., Karasev, D., Forman, W., \aap {\bf 515}, A49 (2010). \\

33. T. Sumi, MNRAS {\bf 349}, 193 (2004). \\

34. A. Udalski, \apj {\bf 590}, 284 (2003). \\
\\
\\

{\it Translated by V. Astakhov}

\end{document}